\documentclass[12pt, preprint]{aastex}

\slugcomment{Not to appear in Nonlearned J., 45.}

 \newcommand{\Msun}{M$_{\odot}$}
\newcommand{\Lsun}{L$_{\odot}$}

\newcommand{\Ox}{[OI]$\lambda$6300}

\newcommand{\Ha}{H$\alpha$}
\newcommand{\SII}{[SII]$\lambda\lambda$6716,6731}

\newcommand{\OI}{[OI]$\lambda\lambda$6300,6363}

\newcommand{\km}{kms$^{-1}$}

\newcommand{\ls}{LS-RCrA 1\ }
\newcommand{\J}{M$_{JUP}$}
\newcommand{\ISO}{ISO-ChaI 217 }
\newcommand{\MASS}{2MASS1207-3932}

\begin{document}


\title{Classical T Tauri-like Outflow Activity in the Brown Dwarf Mass Regime. \altaffilmark{1}}


\author{E.T. Whelan\altaffilmark{2,3}}

\author{T.P. Ray\altaffilmark{2}}

\author{L.Podio\altaffilmark{2}}

\author{F. Bacciotti\altaffilmark{4}}

\author{S. Randich\altaffilmark{4}}

\altaffiltext{1}{Based on data collected by UVES observations (079.C-0375(A)) at the VLT on Cerro Paranal (Chile) which is operated by the European Southern Observatory (ESO).}
\altaffiltext{2}{School of Cosmic Physics, Dublin Institute for Advanced Studies}
\altaffiltext{3}{Laboratoire d'Astrophysique de l'Observatoire de Grenoble}
\altaffiltext{4}{INAF/Osservatorio Astrofisico di Arcetri}


\begin{abstract}

 Over the last number of years spectroscopic studies have strongly supported the assertion that protostellar accretion and outflow activity persists to the lowest masses. Indeed, previous to this work the existence of three brown dwarf (BD) outflows had been confirmed by us. In this paper we present the results of our latest investigation of BD outflow activity and report on the discovery of two new outflows. Observations to date have concentrated on studying the forbidden emission line (FEL) regions of young BDs and in all cases data has been collected using the UV-Visual Echelle Spectrometer (UVES) on the ESO VLT. Offsets in the FEL regions are recovered using spectro-astrometry. Here ISO-Oph 32 is shown to drive a blue-shifted outflow with a radial velocity of 10-20 \km\ and spectro-astrometric analysis constrains the position angle of this outflow to 240$^{\circ}$ $\pm$ 7$^{\circ}$. The BD candidate \ISO is found to have a bipolar outflow bright in several key forbidden lines (V$_{RAD}$ = -20 \km, +40\km) and with a PA of 190$^{\circ}$-210$^{\circ}$. A striking feature of the \ISO outflow is the strong asymmetry between the red and blue-shifted lobes. This asymmetry is revealed in the relative brightness of the two lobes (red-shifted lobe is brighter), the factor of two difference in radial velocity (the red-shifted lobe is faster) and the difference in the electron density (again higher in the red lobe). Such asymmetries are common in jets from low mass protostars (0.1 \Msun to 2 \Msun) and the observation of  a marked asymmetry at such a low mass ( $<$ 0.1 \Msun) supports the idea that BD outflow activity is scaled down from low mass protostellar activity. Also note that although asymmetries are unexceptional, it is uncommon for the red-shifted lobe to be the brightest as some obscuration by the accretion disk is assumed. This phenomenon has only been observed in one other source, the classical T Tauri (CTTS) star RW Aur. The physical mechanism responsible for the brightening of the red-shifted lobe although as yet unknown must also now apply at BD masses to include the \ISO\ outflow. 
 In addition to presenting these new results, a comprehensive comparison is made between BD outflow activity and jets launched by CTTSs.  In particular, the application of current methods for investigating the excitation conditions and mass loss rates ($\dot{M}_{out}$) in CTT jets to BD spectra is explored. Where possible the Bacciotti $\&$ Eisl{\"o}ffel technique is used to study the ionisation fraction, electron temperature and total density. For LS-RCrA 1, \ISO and ISO-Oph 102 $\dot{M}_{out}$ is measured to be in the range 10$^{-10}$ to 10$^{-9}$ \Msun yr$^{-1}$ using a method based on the luminosity of the [OI]$\lambda$6300 and [SII]$\lambda$6731 lines. Mass loss rates for our sample of BD outflows are found to be comparable to the mass accretion rates. 
 Overall, as our results and discussion show, what is currently known about outflow activity in the BD mass regime points to strong similarities between BD outflows and CTT jets.

 \end{abstract}





\section{Introduction}

The outflow phenomenon, much studied in young stars has been observed in a wide variety of astrophysical objects from low mass protostars to active galactic nuclei (AGN) \citep{Reipurth01, Camenzind05}. Brown Dwarfs (BDs) are the newest objects to be added to this list \citep{Fernandez01, Barrado04, Whelan05, Whelan07, Phan08, Whelan09}. We are currently leading a project to search for and study outflows driven by young BDs and very low mass stars. The overall aim of this work is to explore outflow activity in the BD mass regime through comparison with observations of outflows driven by low mass protostars and the classical T Tauri stars (CTTSs) in particular. Much emphasis has been placed on understanding how accretion activity \citep{Gatti06, Herczeg09} and planet formation processes \citep{Pasucci09} differ between low mass protostars and young BDs. However, little is still known about whether the outflow mechanism varies as mass decreases towards the BD regime. Observations aimed at detecting outflows from such low mass objects are difficult and rare and to date six candidates have been investigated by us spectro-scopically at visible wavelengths.

Outflows from low mass protostars are revealed in several forms, from slow wide angled winds to molecular outflows and jets \citep{Reipurth01}. A protostellar jet is a relatively fast (typical radial velocity = 200 \km) collimated outflow normally seen within a few 1000 AU of a young star and bright in optical and/or near-infrared 
shock tracers such as [SII]$\lambda$6731, [FeII] at 1.644$\mu$m or H$_{2}$ at 2.122 $\mu$m. CTTSs are Class II low mass protostars \citep{Lada87, Andre93}, meaning that they are coming towards the end of their accretion/outflow phase and have crossed the so-called ``birth-line" \citep{Stahler83}. As they are optically visible the jet can be traced very close to the protostar, hence, they offer a unique opportunity to study the launching of protostellar jets at high spatial resolution \citep{Ray07}. Jets from CTTSs are traditionally probed at forbidden emission line (FEL) wavelengths and long-slit and integral field spectroscopic techniques have been hugely important in their study \citep{Dougados00, Whelan04}. That CTT-like outflows could be launched by actively accreting BDs was first suggested when high quality spectra revealed the presence of FEL regions in the optical spectra of BDs known to be accretors \citep{Fernandez01}. While the FEL regions were easily identified they were considerably fainter than those detected in the spectra of CTTSs \citep{Masciadri04} and any extension in the form of an outflow was not apparent \citep{Fernandez01, Fernandez05}. Hence their origin in an outflow could not be established. The critical density region for FELs occurs much closer to the driving source of the outflow for nearby ($\sim$ 150 pc) young BDs than for low mass protostars and for BDs is estimated to be on scales of $\sim$ 100 mas \citep{Whelan05}. As the most intense forbidden emission coincides with the critical density region it is currently challenging to directly resolve a BD outflow in FELs and this goal has not yet been achieved. Our approach to this problem has been to obtain high quality spectra using the UV-Visual Echelle Spectrometer (UVES) on the European Southern Observatory's (ESO) Very Large Telescope (VLT) and to recover the spatial offset in the region of forbidden emission using spectro-astrometry.

The sources targeted so far meet at least one of two criteria. Firstly they all have masses at or significantly below the hydrogen burning mass limit (HBML) and secondly they are known to be powerful accretors. Their classification as a strong accretor is based on studies of their \Ha\ lines \citep{Jay03a} and on measurements of their mass accretion rates (see Table 6). In the vast majority of cases the mass of young BD candidates cannot be directly measured and estimates are based on theoretical predictions \citep{Luhman00} and/or spectral modelling \citep{Mohanty04}. 
The objects investigated by us so far are listed in Table \ref{tab1} along with their theoretical masses and spectral types. Previous to this paper we have reported on the discovery of and discussed the optical outflows from, ISO-Oph 102 \citep{Whelan05}, 2MASS1207-3932 \citep{Whelan07} and \ls \citep{Whelan09}. With a mass of only 24 M$_{JUP}$ \MASS\ is at present the lowest mass galactic object driving an outflow. In this paper we present the results of our recent observations of ISO-ChaI 217, DENIS-P J160603.9-205644 and ISO-Oph 32 and report for the first time outflows driven by \ISO and ISO-Oph 32. Of the six objects observed by us to date only one, \ISO\ has a theoretical mass which is just above the HBML, placing it at present just outside the mass range for a BD. However,  with a predicted mass of just 80 \J\ it is still one of the lowest mass objects known to launch an outflow. Therefore for the purpose of the discussion we will include it in this paper with the likely BDs. In addition to presenting our spectro-astrometric analysis, a technique for constraining the position angle (PA) of the BD outflows is outlined, methods for studying the physical conditions of the BD outflow and of estimating the mass outflow rate are discussed and similarities between BD outflows and CTT jets highlighted and considered.

\section{Observations and Spectro-astrometric Analysis}
\label{observations}

Spectra (at orthogonal slit positions) of ISO-Cha1 217, ISO-Oph 32 and DENIS-P J160603.9-205644 were obtained with the high resolution spectrometer UVES (CD3 disperser, 
spectral range 4810-6740 \AA, R=40,000;), on the ESO VLT UT2 \citep{Dekker00}, as part of the observing program 079.C-0375(A). In each case the exposure time was 2700s, the slit width 1\arcsec\ and the seeing varied between 0\farcs5 and 0\farcs9. Spectra were analysed using spectro-astrometry as outlined in previous papers \citep{Whelan05, Whelan07}. Also see \cite{Whelan08} for details on basic data reduction steps and continuum/night sky line subtraction. To summarise spectro-astrometry involves the measurement of the spatial centroid of an emission line (using Gaussian fitting) as a function of velocity. The centroid of the continuum emission is taken as the position of the star/BD and the emission line centroid is measured with respect to this position.

Problems specific to the application of spectro-astrometry to BD spectra are the lack of or weakness of continuum emission from the BD in the range 5000-8000 \AA\ and the faintness of the line emission itself. In order to overcome these problems and increase the spectro-astrometric accuracy, the spectra are summed or smoothed in the dispersion direction. In the absence of detector noise, the spectro-astrometric accuracy is dependent on N$_{p}^{1/2}$ where   N$_{p}$  is the number of detected photons in the line/continuum \citep{Whelan05}. In the cases where spectra are summed, the line and continuum are summed separately and by differing amounts, so that the spectro-astrometric error in the line and continuum position is comparable. Recall that the continuum must be removed before the line centroid is fitted. In these cases, points are a moving sum across the line and continuum (see Figures 3-7 and Figure 11). For BDs where the line emission is weaker than the underlying continuum emission it was more effective to decrease the spectral resolution and hence increase the signal by smoothing the spectrum using a Gaussian filter \citep{Whelan07}. Again this is done so that errors are comparable. The \Ox\ line region in the spectra of ISO-Oph 32 was smoothed using a Gaussian filters of fwhm = 0.2 \AA\ $\times$ 0\farcs32. The results of the spectro-astrometric analysis of the smoothed spectra are shown in Figures  9 and 10. In all cases the dashed line delinates the 1-$\sigma$ error in the continuum (and hence line) centroid. Finally, the issue of spectro-astrometric artifacts must be addressed. Spectro-astrometric artifacts, or false spectro-astrometric signals, can be introduced into a spectrum for a variety of reasons \citep{Podio08}. In our studies of BD outflows we have ruled out the influence of such artifacts by demonstrating that lines like HeI$\lambda$6678, which will not form in an outflow, do not have any spectro-astrometric signal. If the outflow signatures detected in the FELs were due to artifacts then these lines would also show similar offsets. In addition, note that the \Ha\ line in the majority of cases does not show any offset. This is in agreement with the idea that most of the \Ha\ emission comes from the accretion zone close to the source. For all the BDs, spectra are presented at several PAs, and offsets measured at the different PAs agree. Lastly the position velocity (PV) diagram of the emission line regions of \ISO supports the spectro-astrometric analysis.

\begin{table}
\begin{tabular}{lllll}       
 \hline\hline 
 Source                                   &RA (J2000)   &Dec (J2000) & Spectral Type &Mass (\J)     
 \\ 
\hline
ISO-ChaI 217          &11 09 52.0                  &-76 39 12.0  &M6.2                 &80$^{1}$             
\\
\MASS &12 07 33.4 &-39 32 54.0 &M8 &24$^{2}$
\\
DENIS-P J160603.9-205644  &16 06 03.90 &-20 56 44.6 &M7.5 & 40$^{3}$
\\
ISO-Oph 32         &16 26 22.05  &-24 44 37.5 &M8 &40$^{4}$  
 \\
 ISO-Oph 102      &16 27 06.58   &-24 41 47.9  &M6 &60$^{4}$
 \\
\ls   &19 01 33.7  &-37 00 30.0 &M6.5 &35-72$^{5}$
\\  
 \hline  
\end{tabular}
\caption{The spectral type and predicted mass of the BD candidates investigated by us to date. All sources except DENIS-P J160603.9-205644 are found to drive outflows. The numbers 1-6 refer to the papers giving the estimates of the mass of each source, where 1=\cite{Muz05}, 2=\cite{Mohanty07}, 3=\cite{Mohanty04}, 4=\cite{Natta02} and 5=\cite{Barrado04}.
}
\label{tab1}
\end{table}

\section{Results of Spectro-astrometric Analysis}

In this section the results of the spectro-astrometric analysis of the FE and \Ha\ line regions of, ISO-Cha1 217, ISO-Oph 32 and DENIS-P J160603.9-205644, are presented. The particulars of the spectro-astrometric technique, as applied to BD spectra, are discussed above.  As well as recovering the spatial offset in the FEL regions of our BD candidates (FELs were only detected in the case of \ISO and ISO-Oph 32) this technique is also used to estimate the PA of the new BD outflows, through comparison of the spectro-astrometric offsets measured at orthogonal slit positions. Jets driven by CTTSs were first observed using narrow band imagery and long slit spectroscopy \citep{Mundt83, Solf93}. \cite{Hirth94, Hirth97} estimated the micro-jet PA for a sample of CTTSs by taking spectra at several slit PAs and plotting the positional offset and spatial full width half maximum (FWHM) of the [SII]$\lambda$6731 line as a function of slit PA. They argue that the [SII]$\lambda$6731 line is a better indicator of the outflow direction than the [OI]$\lambda$6300 line, as its lower critical density means that it traces the jet at a greater distance from the source. Closer to the source other outflow components e.g the low velocity component could influence any measurement of the jet PA. Although, the approach of \cite{Hirth94} would provide a more straight-forward way of estimating the outflow PA for our BD candidates, the number of spectra that could be collected was limited by the faintness of the FELs and thus the on source time needed to detect them with a reasonable S/N (on average 45 mins with the VLT). For UVES, a negative spectro-astrometric offset, at a N-S slit PA equals a southerly displacement in an emission feature and a positive offset a northerly displacement. For an E-W slit P.A, negative and positive offsets correspond to westerly and easterly directions respectively. This is illustrated in Figure 1. As demonstrated in Figure 2. comparison of the signs of the N-S, E-W spectro-astrometric offsets allows the quadrant in which the blue-shifted outflow lies to be ascertained. Once this is known the actual outflow PA (E of N) can be derived. The dashed, arrowed lines represent the blue-shifted lobe of the outflow as located in each quadrant. The angles A$_{1}$ to A$_{4}$ are simply estimated from {\it tan(A$_{1...4}$) = (offset for E-W slit)/(offset for N-S slit). }

Error is introduced by the fact that the positional offset in the FEL regions is not directly resolved but recovered using spectro-astrometry. The accuracy which one can measure the positional displacement of an unresolved emission feature using Gaussian fitting depends primarily on the S/N of the emission feature. However the displacement is also weighted by the uniformity of the emission. One emission feature could actually be made up of several unresolved features of differing intensity. Where the spectro-astrometric technique measures the spatial mean of the feature to be, is influenced by where the maximum intensity feature lies. In other words an unresolved jet will be measured to be less extended than it really is (using spectro-astrometry), if it actually consists of bright inner knots and lower brightness material further out. For the above reasons the estimates of the PAs of the \ISO and ISO-Oph 32 outflows given here are a first approximation and the errors quoted are taken from the spectro-astrometric uncertainity (dependent on S/N). Also note that outflow PAs measured here only apply to the outflow as studied close to its driving source. It is possible that some bending may occur in the outflow at large distances which will change the outflow PA \citep{Fendt98}. However we expect any such variation to be within the errors quoted in Table 3. Below, the results for ISO-Cha1 217, ISO-Oph 32 and DENIS-P J160603.9-20564 are presented separately.
Tables \ref{tab2} and \ref{tab3} give the FELs detected for each source along with the radial velocity (measured with respect to the systemic velocity), offset measured and estimate of the outflow PA.

\begin{table}
\begin{tabular}{ccccc}       
 \hline\hline 
 Object                                &Slit PA ($^{\circ}$) &Line  &V (\km)  & Offset (mas) 
 \\
 \hline
ISO-ChaI 217                              &0                      &[OI]$\lambda$6300    &-20/30    & -50/50  ($\pm$ 6)
\\
                                                       &                          &[OI]$\lambda$6363       &-10/28  &-50/45 ($\pm$ 9)
\\
                                                        &                         &[SII]$\lambda$6716       &-20/42    &-180/200   ($\pm$ 33.5)                                                     
\\
                                                        &                          &[SII]$\lambda$6731       &-20/38    &-180/180     ($\pm$ 33.5)   
\\
                                                      &90                      &[OI]$\lambda$6300    &-10/20    & -30/25  ($\pm$ 6)
\\
                                                       &                          &[OI]$\lambda$6363       &-10/20  &-40/10  ($\pm$ 12)
\\
                                                        &                         &[SII]$\lambda$6716       &-15/40    &-46/40  ($\pm$ 14)                                                         
\\
                                                        &                          &[SII]$\lambda$6731       &-20/45    &-75/55  ($\pm$ 14)   
\\                                                            
 \hline  

ISO-Oph 32       &0                 &[OI]$\lambda$6300 &-10 &-60 ($\pm$ 10)
\\
                                 &90               &[OI]$\lambda$6300 &-20   &-105 ($\pm$ 19)
\\                                 
 \hline

\end{tabular}
\caption{The FELs of \ISO and ISO-Oph 32 detected at slit PAs of 0$^{\circ}$ and 90$^{\circ}$. Also given are the radial velocity of the lines (blue and red-shifted components in the case of ISO-ChaI 217) and the offsets in milli-arcseconds recovered using spectro-astrometry. An estimate of the outflow PA is made by comparing the offsets measured at the orthogonal PAs.}
\label{tab2}
\end{table}

\begin{table}
\begin{tabular}{ccc}       
 \hline\hline 
 Object                                 &Line    & Outflow PA ($^{\circ}$) 
 \\
 \hline
ISO-ChaI 217                                                   &[OI]$\lambda$6300    & 209 ($\pm$ 8)
\\
                                                                                &[OI]$\lambda$6363       & 206 ($\pm$ 15)
\\
                                                                                &[SII]$\lambda$6716       & 193  ($\pm$ 5)                                                
\\
                                                                                 &[SII]$\lambda$6731       &   200  ($\pm$ 6)
                                                                                 \\
 \hline  

ISO-Oph 32                      &[OI]$\lambda$6300 & 240 ($\pm$ 7)
\\
 \hline

\end{tabular}
\caption{Table giving the position angles of the two outflows observed. The PAs are estimated from orthogonal spectra as described in Section 3. For \ISO a bipolar outflow is uncovered and estimates given above are an average of measurements made for the red and blue-shifted lobes. For ISO-Oph 32 only the blue-shifted outflow is detected.}
\label{tab3}
\end{table}

\begin{figure}
\label{slitpa}
\includegraphics[width=18cm]{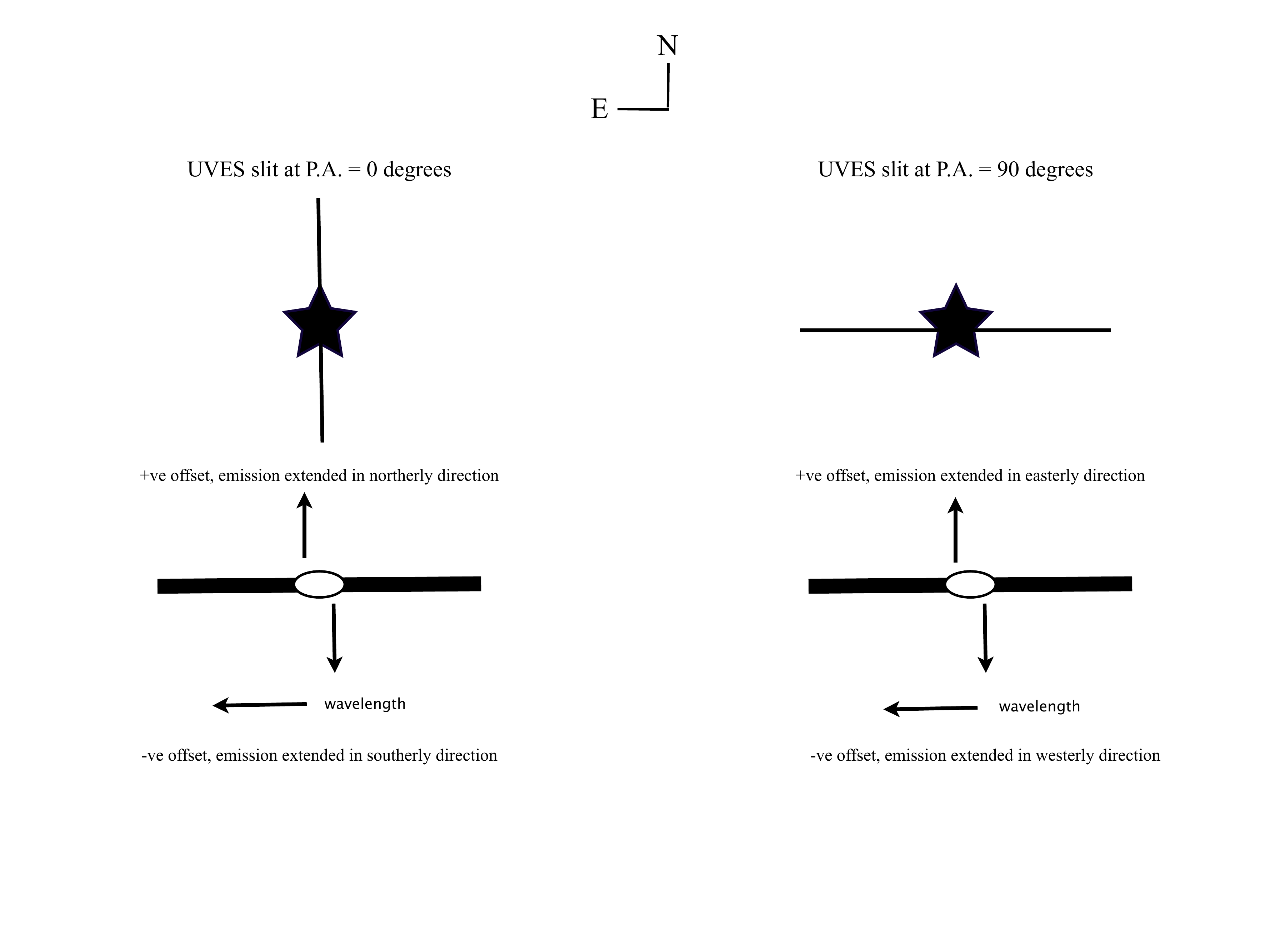}
\caption{Figure illustrating the direction on the sky which the sign of the spectro-astrometric offsets correspond to, for different slit PAs. Above left, the slit is positioned at a PA of 0$^{\circ}$. Below left, the resulting spectrum is shown as a strong emission line feature (assumed to originate in an outflow) superimposed on the continuum emission. A positive spectro-astrometric offset corresponds to extended emission in a northerly direction and a negative offset a southerly direction. In the case of an E-W slit position (PA=90$^{\circ}$) a positive offset corresponds to an easterly direction and negative to a westerly. Wavelength increases towards the left.}
\end{figure}

\begin{figure}
\label{outflowpa}
\includegraphics[width=18cm]{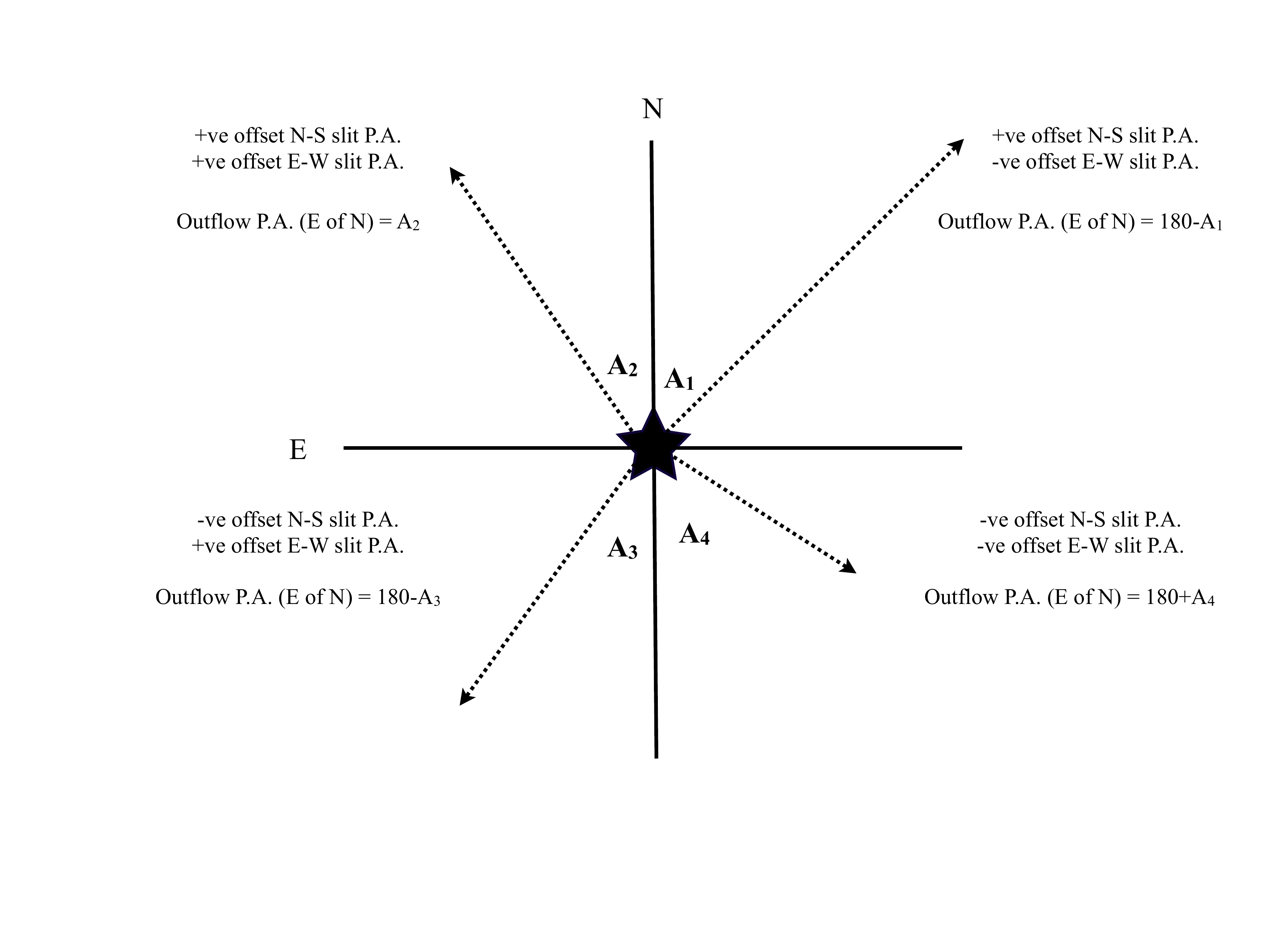}
\caption{Figure illustrating how the signs of the spectro-astrometric offsets can be used to determine the quadrant in which the outflow lies. The dashed lines represent the blue-shifted outflow. Angles A$_{1}$....A$_{4}$ are estimated from the following formula {\it tan(A) = (offset for E-W slit)/(offset for N-S slit). } As negative offsets are measured for both the \ISO and ISO-Oph 32 outflows the blue-shifted lobes lie in the fourth quadrant.}
\end{figure}

\subsection{\ISO}

\ISO was first reported as part of an ISOCAM survey of the Chamaeleon I dark cloud \citep{Persi00}. Current estimates place its mass at 80 \J\ \citep{Muz05} and its spectral type between M7.5 and M6.25 \citep{Lopez04, Muz05}. Hence, it currently lies on the boundary between being a BD or an ultra-low mass star. The accretion properties of \ISO have been much studied in recent years.  \cite{Apai05} confirm mid-infrared excess emission, indicative of the presence of a disk and \cite{Scholz06} demonstrate that it is a strong accretor which exhibits significant variability in its accretion activity. \cite{Scholz06} also observe signatures of outflow activity in their spectra of ISO-Cha1 217. These signatures are in the form of FELs of S. Here the origin of these (and other) FELs in an outflow driven by \ISO is confirmed for the first time.

UVES spectra of ISO-ChaI 217, taken at orthogonal slit positions were obtained on May 1st  2007. Strong \SII, \OI\ and \Ha\ emission was detected at both PAs along with weak [NII]$\lambda$6548 emission. Blue and red-shifted components to the \SII\ and \OI\ lines were detected at an average radial velocity (w.r.t the systemic velocity) of -16 and 32 \km. The systemic velocity of \ISO\ is taken as +2 \km\  \citep{James06, Bary08}. The \SII, \OI\ and \Ha\ lines were analysed with spectro-astrometry and the analysis revealed the blue and red-shifted emission to be offset in opposing directions. Hence, ISO-Cha I 217 is driving a bipolar outflow (see Figures 3, 4, 5, 6, 7). 
As is the case in CTTS the \SII\ lines trace the outflow further from the source than the \OI\ lines (refer to Table \ref{tab2}). Emission is clearly more extended at 0$^{\circ}$ than at 90$^{\circ}$ and a comparison between the two is used to constrain the outflow PA to a range of 193$^{\circ}$ to 208$^{\circ}$ (E of N). Note that, the lines which trace the outflow closest to the source give a slightly larger PA on average than the more displaced line regions (Table 3), although all values agree within the errors. Analysis of the \Ha\ line reveals no detectable contribution from the outflow. Clearly the \Ha\ line is coincident with the continuum position at both position angles. 

A significant feature is the obvious asymmetry in the outflow as traced by [SII]. The red-shifted lobe is much brighter in \SII\ than the blue-shifted and there is a velocity difference of a factor of 2 between the red and the blue lobe. This is especially apparent in the case of the brighter [SII]$\lambda$6731 line at a PA of 0$^{\circ}$ (Figure 7). Although velocity and intensity asymmetries are commonly seen in CTT jets \citep{Hirth94} and while several case have been documented where the red-shifted lobe is the faster it is unusual for the red-shifted lobe to be the brightest. In the majority of CTT jets the circumstellar disk obscures the red-shifted flow. The CTTS RW Aur offers the most striking example of such an asymmetry \citep{Hirth94, Lopez03}. In this case a similar velocity asymmetry has been reported but the slower lobe (red) is the brightest. The \ISO\ asymmetry is clearly visible in the PV diagrams presented in Figure 8. For the purpose of the PV diagrams the [SII]$\lambda$6731 line region was smoothed, using an elliptical Gaussian filter. The PV diagrams support the spectro-astrometric analysis as the opposing offset in the line region at 0$^{\circ}$, measured with spectro-astrometry to be $\sim$ 180 mas, is resolved in the corresponding PV plot. In the PV plot at 90$^{\circ}$ the line appears coincident with the source position again consistent with the much smaller offsets measured at this slit position ($\sim$ 50 mas).

\subsection{DENIS-P J160603.9-205644}

DENIS-P J160603.9-205644 was identified as a BD as part of a survey of the Upper Scorpius OB association \cite{Martin04}. \cite{Mohanty05} classify it as an accretor (\Ha\ 10$\%$ width is given as 306 \km) and give its spectral type and mass as M7.5 and 40 \J. However, no FELs or additonal signatures of outflow activity are found in the spectra of DENIS-P J160603.9-205644. The spectro-astrometric analysis of the \Ha\ line at 0$^{\circ}$ and 90$^{\circ}$ is shown in Figure 11. While the line profile is asymmetric the emission is coincident with the continuum and therefore must originate at the star. The lack of a spectro-astrometric signal in the \Ha\ line rules out any complications with spectro-astrometric artifacts. Although DENIS-P J160603.9-205644 is classified as an accretor it is likely that it is the least accreting object in our sample. \cite{Mohanty05} give upper limits to its mass accretion rate based on the non detection of the Ca II 8662 \AA\ feature. \cite{Herczeg09} measure $\dot{M}_{acc}$ for this object at 2.5 $\times$ 10$^{-12}$ \Msun yr$^{-1}$ consistent with the upper limits set by \cite{Mohanty05}. Also note that \cite{Herczeg09} do detect some faint forbidden emission in the spectrum of DENIS-P J160603.9-205644. Their observations were made with LIRIS on Keck I at a spectral resolution of R=1400. Given the relatively high spectral resolution of our observations and the small accretion rate of this object it is not surprising that no detection is made here.  

\subsection{ISO-Oph 32}

\cite{Natta02} observed this source in the near-infrared and estimated its spectral type and mass at M7.5 and 30-50 \J. They also argue that the mid-infrared excess detected in this object is consistent with the presence of an irradiated disk. In a later study \cite{Natta04} added optical spectra to their near-infrared data. In particular they use the \Ha\ line as a test of accretion activity. In the both the 0$^{\circ}$ and 90$^{\circ}$, high resolution UVES spectrum presented here the \Ox\ line is detected at at S/N of just above 4. As the [OI]$\lambda$6363 line is 3.5 times fainter than the \Ox\ line, it is marginally detected at both PAs and therefore not analysed with spectro-astrometry. The lower critical density lines are below the detection limit in our spectra. 
As shown in Figure 9 analysis reveals the [OI]$\lambda$6300 line to be offset in a southerly direction to $\sim$ 60 mas at 0$^{\circ}$ and $\sim$ 105 mas at 90$^{\circ}$. The average radial velocity (w.r.t the systemic velocity) of the two lines is 15 \km\ and comparison of the measured offsets at the orthogonal slit positions constrains the PA of the blue-shifted outflow at $\sim$ 240$^{\circ}$. The systemic velocity of ISO-Oph 32 is taken as +7 \km\ by \cite{Whelan05}. Again no spectro-astrometric signal is detected in the \Ha\ line, confirming the origin of the bulk of the emission in the accretion flow.

\section{Measuring the Mass Outflow Rate for BD Outflows}

Numerous observational studies have constrained the mass outflow rate ($\dot{M}_{out}$) in CTTSs at $\sim$ 1-10 $\%$ of the mass accretion rate \citep{Hartigan94, Podio06, Rebeca08}. The ratio between the mass outflow and accretion rates ($\dot{M}_{out}$/$\dot{M}_{acc}$) in young stars is a defining characteristic of the mechanism which is at work to launch protostellar outflows \citep{Ferreira06, Ray07}. If a similar ratio between outflow and infall is measured for accreting BDs it will provide compelling evidence that the outflows we are beginning to detect at very low and sub-stellar masses are in fact scaled-down from low mass protostellar outflows. The spectra of low mass protostars contain a wealth of forbidden and permitted lines whose excitation properties are well known. Hence, the ratios and intensities of these lines can be used to probe the physical conditions of the emitting regions. Spectral analysis of known outflow tracers, typically in the optical and near-infrared regimes, has allowed $\dot{M}_{out}$ to be estimated for low mass protostars. In comparison, the spectra of young BDs, have been little investigated. It is only in the last number of years that optical FELs have been detected in BD spectra and comparatively little work has been done in the near-infrared. 

Two distinct methods have been used to measure $\dot{M}_{out}$ in low mass protostars. Here the application of these methods to the problem of measuring mass outflow rates at BD masses, is discussed. The first method (hereafter method A) relies on an estimate of the total density in the outflow $n_{H}$, which in turns requires the electron density (n$_{e}$) and ionisation fraction (x$_{e}$) to be known \citep{Nisini05, Podio06}. These plasma parameters can be calculated using the Bacciotti $\&$ Eisl{\"o}ffel (BE) technique \citep{Bacciotti99, Bacciotti99b}. The BE technique employs selected optical transitions of S$^{+}$, O$^{0}$ and N$^{+}$, to derive x$_{e}$ and T$_{e}$ (electron temperature). The electron density is estimated from the ratio of the [SII] lines, [SII]$\lambda$6716/[SII]$\lambda$6731 \citep{Osterbrock94} and once this quantity is known it is combined with x$_{e}$ to derive n$_{H}$. \cite{Podio06} adopt the BE technique to study the physical conditions in a sample of protostellar outflows and thus calculate their mass loss rate. As the BE technique calculates T$_{e}$ and x$_{e}$ from the [NII]$\lambda\lambda$(6548+6583)/[OI]$\lambda\lambda$(6300+6363) and [SII]$\lambda\lambda$(6716+6731)/[OI]$\lambda\lambda$(6300+6363) line ratios, it requires the use of a given set of elemental abundances. \cite{Podio06} investigate any variations in estimates of  x$_{e}$ and T$_{e}$ for four different abundance sets and calculate  $\dot{M}_{out}$ using, 
 
\begin{equation}
 \dot{M}_{out} =  \mu m_{H} n_{H} \pi r_{J}^{2} v_{J}
 \end{equation}
 
where $\mu$=1.41 is the mean atomic weight, m$_{H}$ the mass of a proton, n$_{H}$ the total density and r$_{J}$, v$_{J}$ the jet radius and velocity respectively.  Here we firstly discuss the analysis of our BD spectra with the BE technique.
 
The BE technique is easily applied to the spectrum of a typical CTTS where the required FELs are well resolved spatially and spectrally with a high S/N. The line flux can be measured using several different line fitting techniques.  The main source of error comes from the uncertainty in the measurement of the line fluxes which depends on the S/N ratio of the line.  In order to use the BE technique to investigate a BD outflow the [OI]$\lambda$6300, [SII]$\lambda$6716, 6731 and [NII]$\lambda$6583 lines must all be detected at a sufficient S/N. The [OI]$\lambda$6363 and [NII]$\lambda$6548 lines can be substituted for the [OI]$\lambda$6300, [NII]$\lambda$6583 lines, if for example the brighter [OI]$\lambda$6300 line is not covered by the spectral range of the instrument or the [NII]$\lambda$6548 is blended with the \Ha\ line. 
Both the [OI]$\lambda$6300 and [NII]$\lambda$6548 lines can be exchanged with the fainter line of their pairs, as the ratio of their intensities is always 
three. This is  due to the fact that they come from the same upper 
level and that the ratio of the A coefficients for spontanoeus emission is three. Of the five BD candidates found so far to have outflows only one, \ls (outflow discussed in Whelan et al. 2009), meets the requirements of the BE technique. For the two lowest mass BDs, ISO-Oph 32 and \MASS\ \citep{Whelan07} only the \OI\ lines are observed. An added difficulty for \MASS\ is that while both the blue and red-shifted lobes of the outflow are detected in the \Ox\ line they are not well resolved spectrally hence the emission from both lobes cannot be properly separated. While both the \OI\ and [SII]$\lambda$6731 lines are found in the spectrum of ISO-Oph 102 \citep{Whelan05} the [SII]$\lambda$6716 line is not detected and therefore there is no estimate of n$_{e}$. Finally for \ISO, while the \Ox\ and \SII\ lines are all observed the [NII]$\lambda$6548 line is not present and the [NII]$\lambda$6583 line is too faint to allow the use of the BE technique. However, using the ratio of the \SII\ lines n$_{e}$ in the blue and red-shifted lobes of the \ISO outflow is measured (see Table \ref{tab4}). 

The results of the application of the BE technique to the blue-shifted outflow driven by \ls are reported in Table \ref{tab4}. 
As described in \cite{Whelan09} UVES spectra of \ls were taken at several P.As between 45$^{\circ}$ and 82$^{\circ}$ and analysis of this data constrained the PA of the \ls outflow at $\sim$ 15$^{\circ}$. Using the BE technique n$_{e}$, x$_{e}$, T$_{e}$ and n$_{H}$ are derived at each PA. The quoted errors come from the uncertainty in the measurement of the line fluxes. 
In the case of low mass protostars, values of x$_{e}$ in the range 0.01 - 0.4, T$_{e}$ in the range 8000-20000 K and n$_{H}$ up to 10$^{6}$ cm$^{-3}$ are typically measured \citep{Ray07}. The estimates of n$_{e}$ for \ls and \ISO are reasonable when compared with estimates made for CTT jets and the large errors (up to $\sim$ 40$\%$) are due to the low S/N in the [SII] lines, which is a low as 3 in some cases. \cite{Comeron03} quoted a value of n$_{e}$ = 2300 cm$^{-3}$ for LS-RCrA 1. Note that the \ISO red-shifted outflow has a higher electron density that the blue-shifted which is a further indication of the asymmetry between the two lobes of the outflow. The low S/N of the \ls [NII]$\lambda$6583 line leads to errors of up to 50$\%$ in the derived values of x$_{e}$ and n$_{H}$. The estimates of x$_{e}$ are very low compared to what one measures for CTTSs and decrease as the slit PA moves away from the outflow PA. In addition, values of T$_{e}$ and n$_{H}$ are very high. With such high densities one would expect more efficient cooling and therefore lower temperatures. The peculiarities in x$_{e}$, T$_{e}$ and n$_{H}$ noted here have been observed in CTT flows where the very beginning of the flow (0\farcs2 to 1\arcsec) is being tested with BE technique. In the case of the BDs we are probing even closer to the driving source of the outflow. \cite{Bacciotti99b} argue that the physical reason for this behaviour could be related to the event that creates the ionisation of the flow e.g. a standing shock. While these preliminary results are interesting, it is impossible at present to say if this approach can reveal anything meaningful about the ionisation fraction / electron temperature and thus shocked emission in BD outflows. Higher S/N spectra of a bigger sample of young BD outflows, aimed at resolving the [NII] lines in particular, should be the next step.

The value of n$_{H}$ estimated from the above method is used in Equation 1 to derive $\dot{M}_{out}$. As the slit PA of 45$^{\circ}$ lies closest to the estimated outflow PA (15$^{\circ}$) n$_{H}$ derived for this slit PA is taken. One of the biggest difficulties with measuring  $\dot{M}_{out}$ for BD outflows is that the outflow width and velocity is not known. Here reasonable estimates of r$_{J}$, v$_{J}$ for \ls are made. \cite{Ray07} show a plot of jet width as a function of projected distance from the source. Assuming a similar relationship for a young BD outflow and considering that offsets of up to 150 mas ($\sim$ 20 AU) have been measured for the \ls FELs, 10 AU is taken as a reasoned approximation of the radius of the \ls outflow. This would suggest an opening angle close to 30$^{\circ}$ for this outflow and is likely an upper limit on the outflow radius. The radial velocities of the \ls FELs are low ($\sim$ 25\km) and at present the inclination angle of the outflow is poorly constrained. Initial observations suggested that the low radial velocity of the FELs and the faintness of the source could be explained by an edge-on accretion disk and thus an outflow inclined close to the plane of the sky. However the lack of red-shifted emission and the evidence for separation between low and high velocity emission in the outflow, presented in \cite{Whelan09}, refute this argument. Assuming that the low radial velocity measured in the outflow tracers suggests a low outflow inclination angle (20-30$^{\circ}$), for the purpose of calculating $\dot{M}_{out}$, v$_{J}$ is taken at 75 \km. Combining n$_{H}$ with r$_{J}$ and v$_{J}$ yields $\dot{M}_{out}$ = 2.4 $\times$ 10$^{-9}$ \Msun yr$^{-1}$.

The second widely used method for measuring $\dot{M}_{out}$ (hereafter method B) is based on the observed luminosity ${\it L(line)}$ of an optically thin line such as [SII] or [OI]. The luminosity of a specific line 
can be derived as follows \citep{Comeron03}
\begin{equation}
L(L_{\odot}) = 6.71 \times 10^{-5}D^{2}(pc)EW(\AA) 10^{-0.4R_{0}}
\end{equation}
(D is the distance to the object, EW the equivalent width of the line and R$_{0}$ the dereddened magnitude) and used to evaluate the mass of the flow ${\it M}$ (in the aperture) according to the following equations \citep{Hartigan95}, 
\begin{equation}
M = 9.61 \times 10^{-6} (1 + \frac{n_{c}}{n_{e}}) ( \frac{L_{6300}}{L_{\odot}}) M_{\odot}
\end{equation}
\begin{equation}
M = 1.43 \times 10^{-3} (1 + \frac{n_{c}}{n_{e}})  ( \frac{L_{6731}}{L_{\odot}}) M_{\odot}
\end{equation}
$\dot{M}_{out}$ = MV$_{tan}$/l$_{tan}$ where n$_{c}$, V$_{tan}$ and l$_{tan}$ are the critical density, outflow tangential velocity and the size of the aperture in the plane of the sky. As n$_{e}$ for CTTSs is comparable to the critcal density of [SII]$\lambda$6731 \cite{Hartigan95} remove the critical density dependence from Equation 4 above. For the BD outflows investigated to date n$_{e}$ $<$ n$_{c}$ hence this dependence is included here.

In order to measure ${\it L(line)}$ and apply the above method to our sources, R$_{0}$, n$_{e}$ and V$_{tan}$ must be known. Here we estimate $\dot{M}_{out}$ for \ls, \ISO and ISO-Oph 102. The values of EW, R$_{0}$, n$_{e}$ and V$_{tan}$ used in each case are given in Table \ref{tab5}. For \ls $\dot{M}_{out}$ is measured at 6.1 $\times$ 10$^{-10}$ \Msun yr$^{-1}$ and 2.0 $\times$ 10$^{-10}$ \Msun yr$^{-1}$ from the [OI]$\lambda$6300 and [SII]$\lambda$6731 lines respectively. V$_{tan}$ is taken at -43 \km. This estimate is based on a radial velocity of 25 \km\ and an outflow inclination angle of 30$^{\circ}$ (see above). For the \ISO bipolar outflow we chose to estimate $\dot{M}_{out}$ from the [SII]$\lambda$6731 line luminosity.
As the outflow is resolved spectrally in the [SII]$\lambda$6731 line (the lobes are blended spectrally in the [OI]$\lambda$6300 line) it was possible to estimate $\dot{M}_{out}$ in both lobes of the outflow. For the faster (red-shifted lobe)  $\dot{M}_{out}$ =  3.1 $\times$ 10$^{-10}$ \Msun yr$^{-1}$. This is almost twice the outflow rate of the blue lobe, which is measured at 1.8 $\times$ 10$^{-10}$ \Msun yr$^{-1}$. A similar difference in $\dot{M}_{out}$ is observed in the two lobes of the RW Aur asymmetric jet \cite{Woitas02}. R$_{0}$ for \ISO is taken from the R = 19.59 mag \citep{Lopez04} and A$_{J}$ given by \cite{Luhman04} as 0.7 mag. As we see the red-shifted component  of the \ISO outflow the outflow likely has a low inclination angle. Radial velocities of $\sim$ -20 \km\ and 40 \km\ are measured for the red and blue lobes therefore tangential velocities of -55 \km\ and 110 \km\ are assumed for the purpose of this calculation. This is based on an outflow inclination angle of 20$^{\circ}$.

Finally we also apply the method of \cite{Hartigan95} to the blue-shifted outflow driven by ISO-Oph 102. The [SII]$\lambda$6731 line is detected at a low S/N and its luminosity is estimated at 3.1 $\times$ 10$^{-7}$ \Lsun. As the [SII]$\lambda$6716 line is not detected there is no estimate of n$_{e}$ for the outflow. However, as n$_{c}$ for the [SII]$\lambda$6731 line is closer to the value of n$_{e}$ measured for CTT / BD outflows (than for the [OI]$\lambda$6300 line), Equation 4 is less sensitive to the n$_{c}$ / n$_{e}$ dependence and we can make a reasonable estimate of the possible range of n$_{e}$ for the ISO-Oph 102 outflow. Using a range of 10$^{3}$ cm$^{-3}$ to 10$^{4}$ cm$^{-3}$ for n$_{e}$ gives $\dot{M}_{out}$ at 1.7-11.8$\times$ 10$^{-10}$ \Msun yr$^{-1}$. A value of V$_{tan}$ = 78 \km\ is assumed for this outflow based on the measured radial velocity of its FELs \citep{Whelan05} and on a outflow inclination of 30$^{\circ}$. \cite{Phan08} estimate the inclination angle of the ISO-Oph 102 disk at  63$^{\circ}$ to 66$^{\circ}$. This suggests that the outflow inclination lies close to 30$^{\circ}$ as assumed. R$_{0}$ is taken from \cite{Natta04} where A$_{v}$ is estimated at 3 mag. Note that \cite{Phan08} measure $\dot{M}_{out}$ = 1.4 $\times$ 10$^{-9}$  \Msun yr$^{-1}$ for the ISO-Oph 102 molecular outflow. It is reasonable that  $\dot{M}_{out}$ for a molecular outflow be greater than the outflow rate in the underlying jet. Assuming that the jet is powering the molecular outflow \citep{Downes07}, the mass outflow rate of the molecular component will grow with time as the jet transfers increasing amounts of energy and momentum.

Table 6. compares $\dot{M}_{out}$ and $\dot{M}_{acc}$  for LS-RCrA 1, \ISO and ISO-Oph 102. As the typical width of a BD outflow is not known but reasonable estimates of outflow velocity can be made from measured radial velocities, method B is more suitable at present as a tool for estimating  $\dot{M}_{out}$. An added advantage of method B is that Equation 4 is less sensitive to n$_{e}$ which is useful for cases where n$_{e}$ is not known and the [SII]$\lambda$6731 line alone is detected. Method B has allowed us to limit $\dot{M}_{out}$ for all three objects to the range 10$^{-10}$ to 10$^{-9}$ \Msun yr$^{-1}$. Quantities such as jet velocity and electron density must be better constrained in order to tighten this range. While the measured range for  $\dot{M}_{acc}$ is at present equally as large and is of order a few times 10$^{-10}$ \Msun yr$^{-1}$  up to of order 10$^{-9}$, what can be concluded here is that $\dot{M}_{out}$ for the BDs in our sample is {\it comparable} to $\dot{M}_{acc}$. 
However, some important points to note are as follows. The number of objects in our sample is very small and the standard $\dot{M}_{acc}$ versus $\dot{M}_{out}$ correlation for CTTSs shows a very wide scatter either side of the fitted least squares straight line \citep{Hartigan95}. Importantly, there may also be some selection effects in our sample. As our observations are difficult and at the edge of what the VLT can do (the [OI]$\lambda$6300 line emission is only just observable), one would therefore tend to pick up only objects with higher outflow rates (brighter [OI] emission) all else 
being equal. In other words, in the plot (for CTTSs) mentioned above we might be 
selecting only the brighter (upper) points in the scatter.

Difficulties related to measuring $\dot{M}_{acc}$ in BDs should also be considered. The most common methods for deriving $\dot{M}_{acc}$ are based on modelling of lines like \Ha\ or Pa$\beta$ \citep{Natta04}, on veiling measurements and on the flux of CaII lines like the CaII$\lambda$8662 \citep{Mohanty05}. However accretion rate estimates for BDs based on different lines (e.g. Halpha or Ca H$\&$K) give very different results \citep{Mohanty05}.
Individual lines (e.g. H$\alpha$) vary on time-scales of weeks \citep{Scholz05, Nguyen09} and thus 
the rates measured at a particular epoch may not reflect the long term 
accretion rate that is more appropriate for comparison with  $\dot{M}_{out}$. A second concieveable difficulty with these approaches is posed by any possible contribution to the \Ha\ or CaII lines from an outflow. For example  \cite{Whelan09} measure a significant contribution to the \ls\ outflow in the wings of the \Ha\ line. Overall, further investigation is needed to confirm if in BDs a larger fraction of infalling material is re-ejected in an outflow than in low mass stars. One theory of BD formation is that they originate from cores that never accrete enough material to enter a H burning phase. A mass outflow rate which is a significant fraction of the infall rate could account for this. Future work on measuring $\dot{M}_{out}$ for young BD outflows should focus on increasing the number of objects known to have outflows, on obtaining better quality spectra and on imaging studies to allow outflow width and velocity to be derived. This work would also benefit from expansion into the NIR regime \citep{Podio06}.

\begin{table}
\begin{tabular}{llllll}       
 \hline\hline 
 Object                        & Slit PA ($^{\circ}$)  &n$_{e}$ (cm$^{-3}$)  &x$_{e}$  &T$_{e}$ (10$^{3}$ K) &n$_{H}$ (10$^{5}$ cm$^{-3}$)
 \\
 \hline
\ls                                & 45                                      &1478 ($\pm$ 170) &0.012 ($\pm$ 0.005)  &35 ($\pm$ 1) &1.2 ($\pm$ 0.7)
\\
                                    & 63                                      &1300 ($\pm$ 140) &0.004 ($\pm$ 0.002)  &37 ($\pm$ 2) &3.3 ($\pm$ 2.0)
\\    
                                    & 74                                      &1300 ($\pm$ 120)  &0.002 ($\pm$ 0.001) &41 ($\pm$ 1) &6.7 ($\pm$ 2.3)
                                    \\
                                    & 82                                      &1500 ($\pm$ 200) &0.002 ($\pm$ 0.001)  &44 ($\pm$  3) &8.6 ($\pm$ 2.7)
\\                                                   
 \hline  
\ISO     
                      &0 B &1900 ($\pm$ 700) &-- &-- &--
\\
                      &0 R &5700 ($\pm$ 2300) &-- &-- &--
\\
                      &90 B  & 600 ($\pm$ 200) &-- &-- &--
\\
                      &90 R    &2100 ($\pm$ 600) &-- &-- &--
\\                                                              
 \hline

\end{tabular}
\caption{Physical conditions in the outflows of \ls\ and \ISO\ as defined by n$_{e}$, x$_{e}$, T$_{e}$ and n$_{H}$. The electron density n$_{e}$ is estimated from the ratio of the [SII]$\lambda$6716/[SII]$\lambda$6731 lines. As explained above x$_{e}$, T$_{e}$ and n$_{H}$ could only be measured in the case of \ls and this was done using the BE technique. While estimates of n$_{e}$ are reasonable when compared to  measurements made for low mass protostellar outflows, values of x$_{e}$ are far less and T$_{e}$, n$_{H}$ far greater than would be expected. Several factors make it currently very difficult to applying techniques such as the BE technique to BD spectra (refer to section 4.4)}
\label{tab4}
\end{table}

\begin{table}
\begin{tabular}{lllll}       
 \hline\hline 
                                                             &\ls                                               &\ISO (red)                          &\ISO (blue)                                            &ISO-Oph 102 
 \\
 \hline
EW [OI]$\lambda$6300 (\AA)         &-30.0                                         & -                                         & -                                                              &-2
\\
EW [SII]$\lambda$6731 (\AA)        &-6.4                                            & -0.7                                   &-0.3                                                         &-0.2
\\    
 R$_{0}$                                             &19.2$^{*}$                                &16.59                                &16.59                                                       &14.57
\\
L [OI]$\lambda$6300 (\Lsun)         &7.1 $\times$ 10$^{-7}$          & -                                         & -                                                               &3.1 $\times$ 10$^{-6}$
\\   
L [SII]$\lambda$6731 (\Lsun)        &1.4 $\times$ 10$^{-7}$          &2.1 $\times$ 10$^{-7}$  &9.1 $\times$ 10$^{-8}$                        &3.1 $\times$ 10$^{-7}$

\\
n$_{e}$ (cm$^{-3}$)                        &1300                                        &3900                                   &1250                                                        &10$^{3}$ -10$^{4}$
\\
i ($^{\circ}$)                                       &30                                             &20                                        &20                                                           &30
\\
V$_{tan}$ (\km)                                &43                                             &55                                         &110                                                         &78
\\
D (pc)                                                 &130                                          &140                                      &140                                                          &125
\\
l$_{tan}$ (\arcsec)                           &1.2                                           &1                                             &1                                                             &1
\\                                               
                                               \hline  
\end{tabular}
\caption{Values used in the calculation of $\dot{M}_{out}$ from the method of \cite{Hartigan95}. As both lobes of the \ISO outflow are not resolved spectrally in the [OI]$\lambda$6300 line it was not possible to measure their separate EWs so no values are given here. * R$_{0}$ is the de-reddened magnitude and for \ls\ is taken from \cite{Comeron03}.}
\label{tab5}
\end{table}

 \begin{table}
\begin{tabular}{ccc}       
 \hline\hline 
 Object                                 &$\dot{M}_{out}$ (\Msun yr$^{-1}$)    & Method 
 \\
 \hline
\ls                                                  & & 
\\
                                                      &2.4 $\times$ 10$^{-9}$       &A                                          
\\
                                                      &6.1 $\times$ 10$^{-10}$     &B [OI]$\lambda$6300       
\\
                                                      &2.0 $\times$ 10$^{-10}$     &B [SII]$\lambda$6731                                                 
\\
 ISO-Oph 102                         & &                     
 \\
                                                       &1.7-11.8 $\times$ 10$^{-10}$        &B [SII]$\lambda$6731        
\\
                                                       &1.4 $\times$ 10$^{-9}$                    &CO molecular outflow$^{1}$
\\                                                       
\ISO Red flow
\\
                                                      &3.1 $\times$ 10$^{-10}$   &B [SII]$\lambda$6731
\\
\ISO Blue flow 
\\                                                     
                                             &1.8 $\times$ 10$^{-10}$   &B [SII]$\lambda$6731
\\
\hline

\end{tabular}

\vspace{1cm}

\begin{tabular}{ccc}       
 \hline\hline 
 Object                                 &$\dot{M}_{acc}$ (\Msun yr$^{-1}$)    & Method 
 \\
 \hline
\ls                                                  & & 
\\
                                                      &2.8 $\times$ 10$^{-10}$       &CaII($\lambda$8542)$^{2}$                                         
\\
                                                      &10$^{-10}$-10$^{-9}$           &Optical Veiling$^{3}$     
\\
                                                      &10$^{-9}$                                &CaII($\lambda$8662)$^{4}$                                                  
\\
                                                      &10$^{-10}$                              &\Ha\ 10$\%$ width$^{5}$
\\  
 ISO-Oph 102                         & &                     
 \\
                                                       &10$^{-9}$        &\Ha\ 10$\%$ width$^{6}$         
\\
                                                       &4.3 $\times$ 10$^{-10}$  & J and K band spectra$^{7}$
                                                       \\                                                       
\ISO
\\
                                                      &1.0 $\times$ 10$^{-10}$ &\Ha\ emission$^{8}$ 
\\
 \hline

\end{tabular}
\label{tab6}
\caption{Measurements of $\dot{M}_{out}$ and $\dot{M}_{acc}$ for the objects studied to date. For the BDs in our sample $\dot{M}_{out}$ is comparable to $\dot{M}_{acc}$. 1=\cite{Phan08}, 2=\cite{Comeron03}, 3=\cite{Barrado04}, 4=\cite{Mohanty05}, 5=\cite{Scholz06}, 6=\cite{Natta04}, 7=\cite{Natta06}, 8=\cite{Muz05}}
\end{table}

\section{Classical T Tauri-like Outflow Activity in the Brown Dwarf Mass Regime}
\label{discussion}

Much of what is now known about the launching of protostellar jets comes from investigating the FEL regions of CTTSs. The first long-slit spectroscopic studies allowed their morphology, kinematics and position angle to be studied \citep{Hirth97} and progress in this area has been driven by high angular resolution techniques \citep{Ray07}. A notable feature of the FEL regions of CTTSs is that they often have two velocity components, a low velocity component (LVC) believed to trace a wide disk wind and a high velocity component (HVC) tracing the jet. Currently, the scale over which jets are collimated, the jet rotation, the jet excitation conditions and mass outflow rate are all estimated from their FEL regions \citep{Ray07}. While the distinctive FEL profiles associated with outflow activity were first studied in the spectra of CTTSs \citep{Hirth97}, Class I low mass protostars and Herbig Ae/Be stars are now also known to show strong FE. Studies have compared the FEL regions of the above objects to those of the CTTSs \citep{Bohm94, Corcoran98, Davis03, Rebeca08}. The earliest studies which identified FEL tracers of mass outflow in the spectra of BDs include \cite{Fernandez01} and \cite{Barrado04}. \cite{Whelan07} presented evidence which supported the hypothesis that BD outflows are scaled-down from CTT outflows. Here recent new evidence is summarised. As all the BDs investigated to date are optically visible it is assumed that they are the substellar mass counterparts to the CTTSs i.e. equivalent to Class II low mass sources. 
 
For the five BDs whose FEL spectra have been investigated to date the [OI]$\lambda$6300 line is the dominant line. The [SII]$\lambda$6731 line is detected in three cases (LS-RCrA 1, ISO-ChaI 217 and ISO-Oph 102) and only for \ls is the [NII]$\lambda$6583 line detected at a reasonable S/N. Very faint [NII]$\lambda$6583 emission is found in the spectrum of ISO-ChaI 217.  Interestingly for \ls \cite{Whelan09} detected both the blue and red-shifted outflow in the wings of the \Ha\ line. This was the first time \Ha\ was shown to originate in a BD outflow. As only blue-shifted FE is found this points to a dust hole in the \ls disk, of an estimated size of $\sim$ 7AU (refer to Whelan et al 2009 for further information).  Spectro-astrometric studies of permitted emission lines have led to similar discoveries for the disk of CTTSs \citep{Whelan04, Takami01}. This is an important parallel between the disk-outflow systems of BDs and CTTSs. 

A second striking parallel between outflow activity in BDs and CTTSs is offered by the emergence of evidence of a LVC to BD FELs. Again this comes from the recent study of LS-RCrA 1.  Although the LVC and HVC are not resolved spectrally for the \ls FELs, \cite{Whelan09} argue that the large spread in radial velocity with the [OI]$\lambda$6300 line having the lowest velocity (-5 \km) and the [NII]$\lambda$6583 line the highest (-22 \km\ with a wing extending to -75 \km) makes it clear that the [OI]$\lambda$6300 line is predominantly tracing a LVC component to the outflow and the [NII]$\lambda$6583 line the HVC. The [SII]$\lambda$6731 line shows evidence of both a LVC and a HVC. In the case of ISO-Oph 32 the low radial velocity of the [OI]$\lambda$6300 emission line is also suggestive of the presence of a LVC to the outflow driven by this BD. An explanation for the low velocity based on a edge-on disk scenario is less likely given the lack of red-shifted emission (the \MASS\ FELs also have low radial velocities but both blue and red-shifted emission is seen) hence it is likely that the [OI]$\lambda$6300 line has a strong contribution from a LVC. The confirmation of a LVC to BD outflows  would makes it very likely that many of the same processes active in the launching region of BDs and CTTSs and thus strongly links BD and protostellar outflows. However a large sample of young BDs must be investigated to obtain this confirmation. 

As asymmetries are frequently observed in CTT outflows a third compelling piece of evidence supporting the continuation of CTT-like outflow activity into the BD regime, is the asymmetrical nature of the \ISO outflow. \cite{Hirth94} first reported asymmetries in the jets driven by the CTTS RW Aur and DO Tau. For RW Aur the brighter red-shifted lobe has a radial velocity a factor of  2 slower than the blue-shifted lobe. In the case of DO Tau, while the blue-shifted lobe is the brightest it is slower than the red-shifted flow by a factor of 2.3. In addition, the results of \cite{Hirth94} reveal a difference in the electron density between both lobes of the flows. For RW Aur n$_{e}$ is higher in the red lobe than the blue, however, for DO Tau the brighter lobe has the lowest electron density. \cite{Hirth94} also list several other protostars for whom radial velocity asymmetries have been reported highlighting that this is a common feature of protostellar outflows. The discovery of such strong asymmetries so close to the driving source is important as it shows that these asymmetries are intrinsic to the central engine and suggests that their origin is in the jet launching mechanism itself. \cite{Ferreira06} also discuss the origin of asymmetries in CTT jets. The asymmetries in the FELs of \ISO\ are remarkably similar to what has been well studied for RW Aur. In the spectra taken at both 0$^{\circ}$ and 90$^{\circ}$ the radial velocity of the red-shifted component is a factor of $\sim$ 2 higher than that of the blue-shifted lobe, the red-shifted lobe is brighter and initial results suggest that the electron density of the red-shifted flow is also higher. Spectra taken along the actual outflow PA (estimated here at $\sim$ 200$^{\circ}$) would allow a more detailed investigation of the \ISO\ outflow asymmetries. 

Finally we consider the existing evidence that BD outflows also have a molecular component. Although early observations of CTT jets concentrated on the optical regime observations now extend into the near-infrared \citep{Davis03, Beck08}. \cite{Takami04} first reported a detection of a warm molecular component to the DG Tau jet (observed in the H$_{2}$, v=1-0 S(1) line) and observations of H$_{2}$ in the spectra of CTTSs are now not unusual. Note however that shocked H$_{2}$ emission in outflows is much more common where a source is in the embedded phase, i.e. Class 0 - I \citep{Davis01}. While there has been very little investigation of BD outflow activity in the NIR, \ls is known to show strong H$_{2}$ emission at 2.122 $\mu$m \citep{Fernandez01}. A recent significant discovery in the field of BD outflows is the observation of a CO J = 2 -1 molecular outflow driven by ISO-Oph 102 \citep{Phan08}. ISO-Oph 102 was the first BD shown by us to drive an outflow \citep{Whelan05}. Blue-shifted FE was detected in its optical spectrum and found to be offset in a southerly direction. \cite{Phan08} used the sub-millimeter array (SMA) to  observe a small-scale molecular outflow driven by ISO-Oph 102. The total scale of the outflow is $\sim$ 20\arcsec\ and it is aligned at a N-S PA. CTTSs are optically visible and therefore no longer embedded in their natal material. It is argued that molecular outflows originate as circumstellar material is entrained and driven outwards by an underlying jet \citep{Downes07}, therefore the most spectacular protostellar molecular outflows are driven by embedded protostars such as Class 0 or Class I low mass stars \citep{Arce07}. CO outflows are driven by the youngest CTTSs \citep{Launhardt09} and observations are rare. The existence of a molecular component to BD outflows is a fourth noteworthy similarity between BD and low mass protostellar outflows.

\section{Summary}

Currently our understanding of astrophysical outflows is focussed on observations of low mass protostellar objects. Although much can be gleaned from such observations a proper appreciation of the outflow phenomenon can only be reached through a complete investigation of how outflow activity compares for a range of driving sources, varying in mass and age. The new results included in this paper include the presentation of two new BD outflows, the unusual red-shifted asymmetry in the \ISO\ and the demonstration that  $\dot{M}_{out}$ in BDs is comparable to $\dot{M}_{acc}$. Overall, the main aim of our BD project is to explore how substellar outflow activity compares to what is well documented for low mass protostars. As our sources are all optically visible our comparison is focussed on the CTTSs. A second advantage of this work is that it may add to the debate on the formation mechanism of BDs i.e. whether they form like low mass protostars or are ejected embryos from multiple systems. The later mechanism is thought to produce truncated disks and very little envelope. If BD outflows are long-lived this would argue against such a scenario although further investigation is required. 

Due to the faintness of the BD outflows observations are difficult and time consuming. However data gathered to date has led to some significant results. The most compelling pieces of evidence which support the continuation of CTT-like outflow activity into the BD mass regime include, the detection of an outflow component to the \ls \Ha\ emission line and the suggestion of a possible disk dust hole, the identification of an LVC to the FELs of \ls and ISO-Oph 32, the strong asymmetry in the \ISO outflow and the discovery of a molecular component to the optical outflows of \ls and ISO-Oph 102.  

While these results are significant much further work is needed to provide a more complete picture of how BD and CTT outflows compare. This is especially true for any study of the physical conditions in BD outflows and BD mass outflow rates. A considerably increased sample is needed. This can be provided through further optical and NIR spectroscopic studies complemented by millimetre studies.  Investigations must be extended to other wavelength regimes so that NIR atomic and molecular outflow tracers can be identified and further CO outflows found. Finally optical/NIR images would allow the collimation of the BD outflows to be probed. Many of these projects are under-way or planned however due to faintness of the emission from a typical BD outflows, it is likely that long-slit/echelle spectroscopy combined with spectro-astrometry will remain (in the short-term at least) the most effective way to investigate BD outflow activity. 

\begin{center}
\begin{figure}
\includegraphics[width=9.5cm]{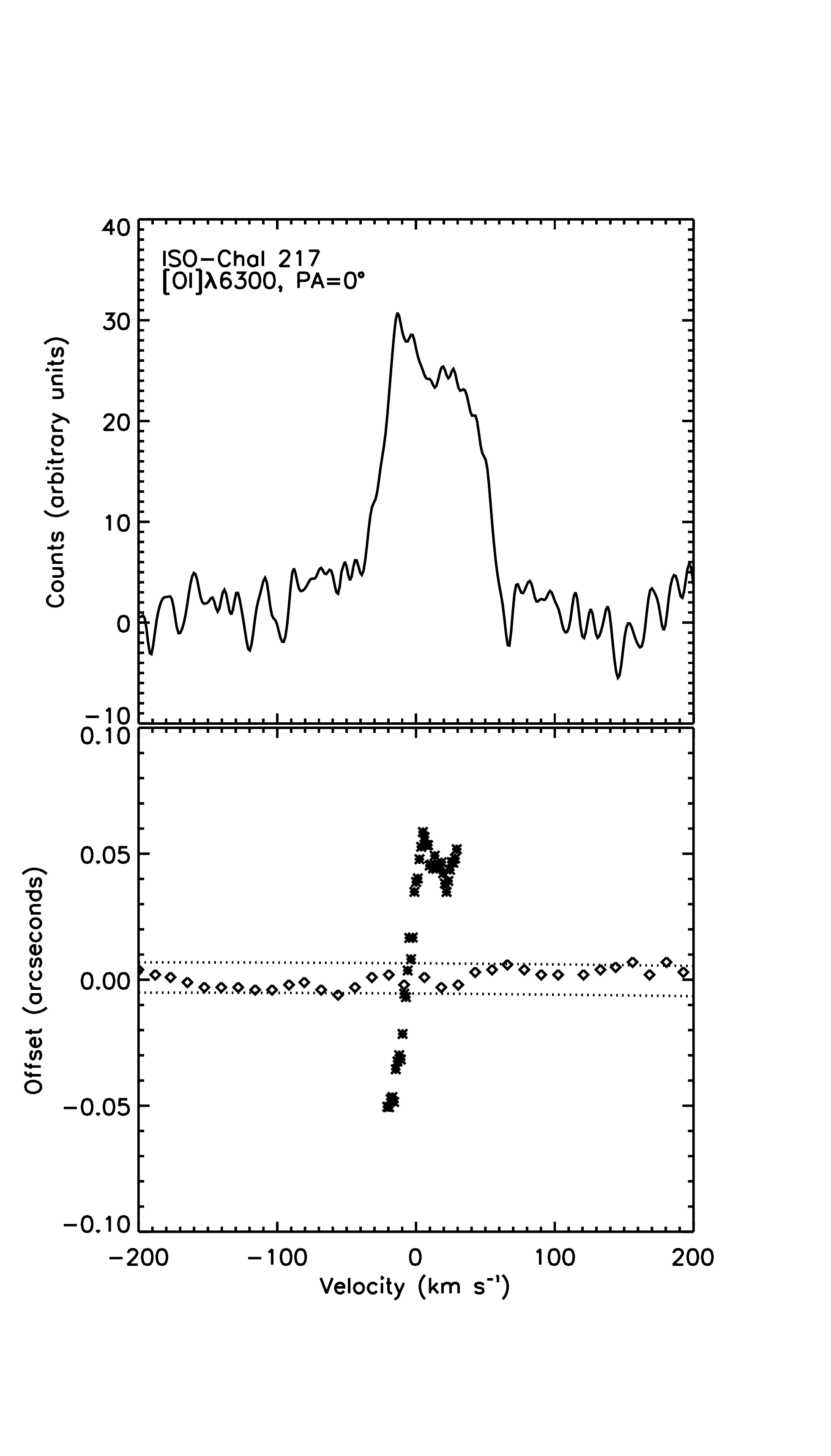}
\includegraphics[width=9.5cm]{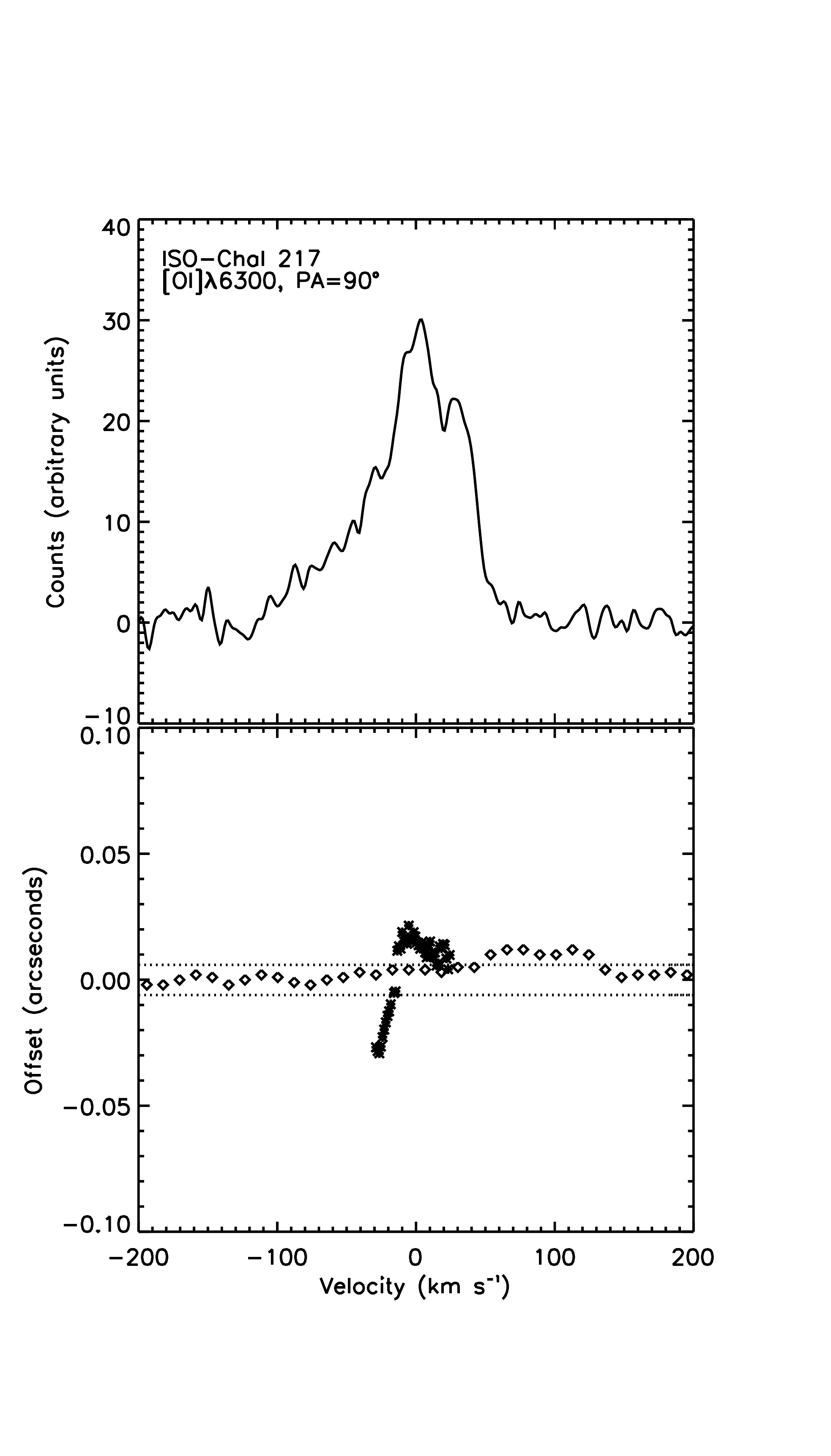}
 \caption{Spectro-astrometric analysis of the ISO-Cha217 \Ox\ line. The line has a blue and a red-shifted component. In the spectrum taken at 90$^{\circ}$ the blue-shifted component to the line appears as an extended wing. Offsets decrease at 90$^{\circ}$ and comparison with measurements at 0$^{\circ}$ constrain the outflow PA at $\sim$ 208$^{\circ}$. Offsets are a moving sum across the line and continuum position. The displacement in the line is measured in the continuum subtracted spectrum. The 1-$\sigma$ error in the line is marked by the dashed line. Finally as outlined in section 2 the line and continuum emission is summed in such a way that the 1-$\sigma$ error in both is comparable.}
 \end{figure}
 \end{center}

 \begin{center}
\begin{figure}
\includegraphics[width=9.5cm]{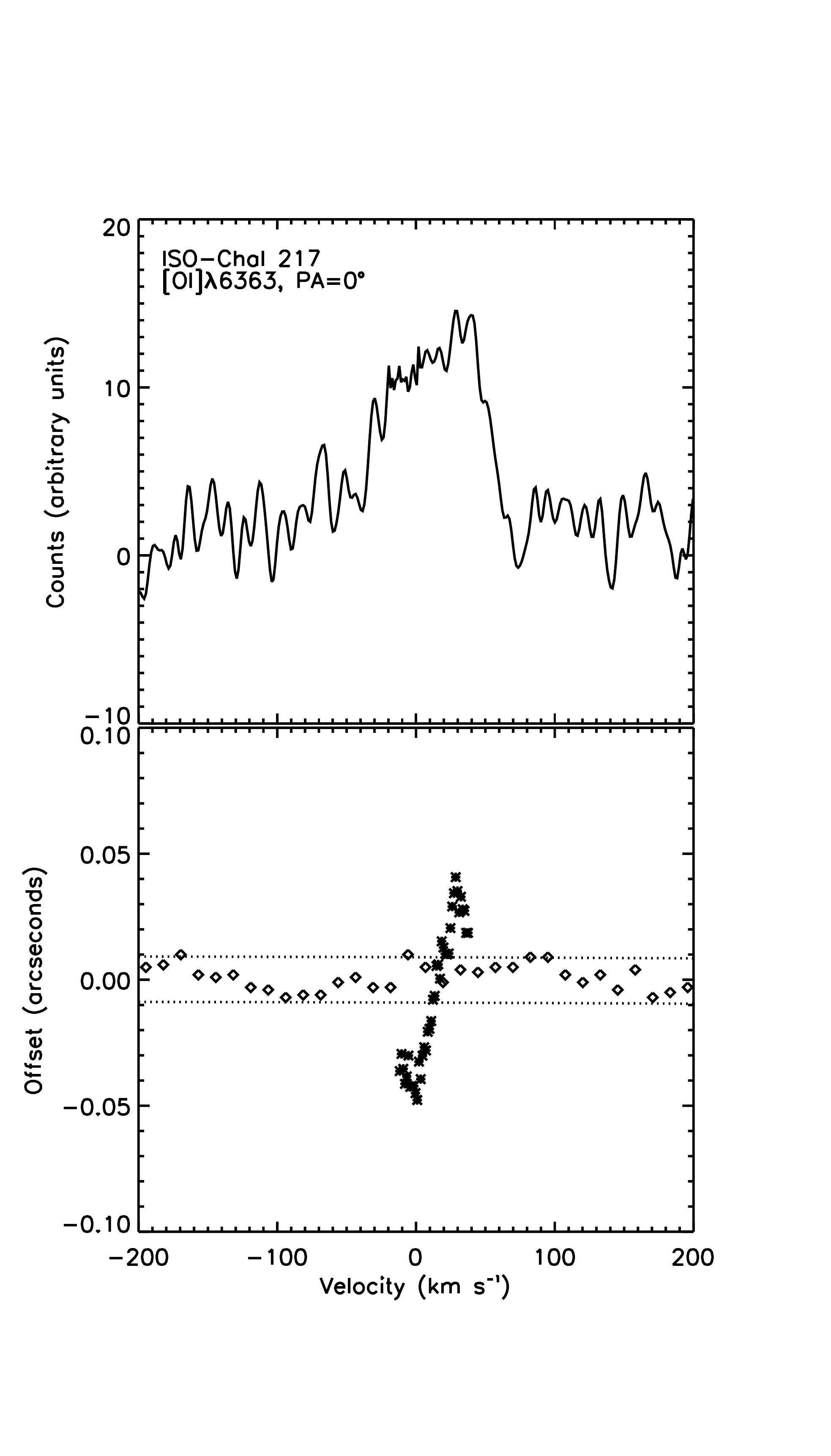}
\includegraphics[width=9.5cm]{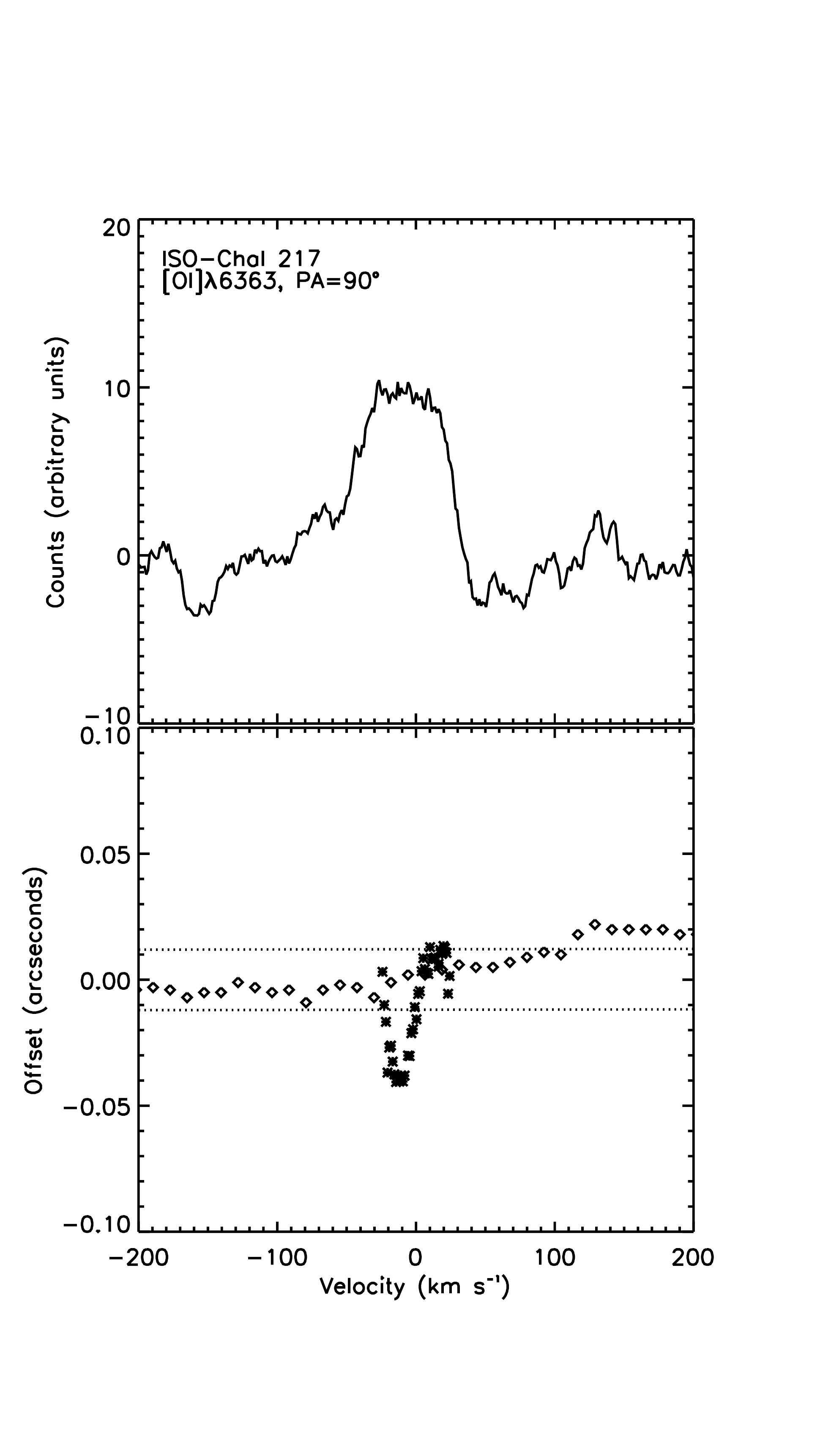}
 \caption{Spectro-astrometric analysis of the ISO-Cha217 [OI]$\lambda$6363 line. Analogous to the \Ox\ line, the line is made of blue and red-shifted emission which spectro-astrometry shows to be offset in opposing directions. As expected the offsets are similar to what is measured for the \Ox\ line. Again comparison between measurements at 0$^{\circ}$ and 90$^{\circ}$ constrain the outflow PA at $\sim$ 205$^{\circ}$. As for all the ISO-ChaI 217 FELs, offsets are a moving sum across the continuum and line position and the 1-$\sigma$ error in both measurements is comparable. }
 \end{figure}
 \end{center}
 
  \begin{center}
\begin{figure}
\includegraphics[width=9.5cm]{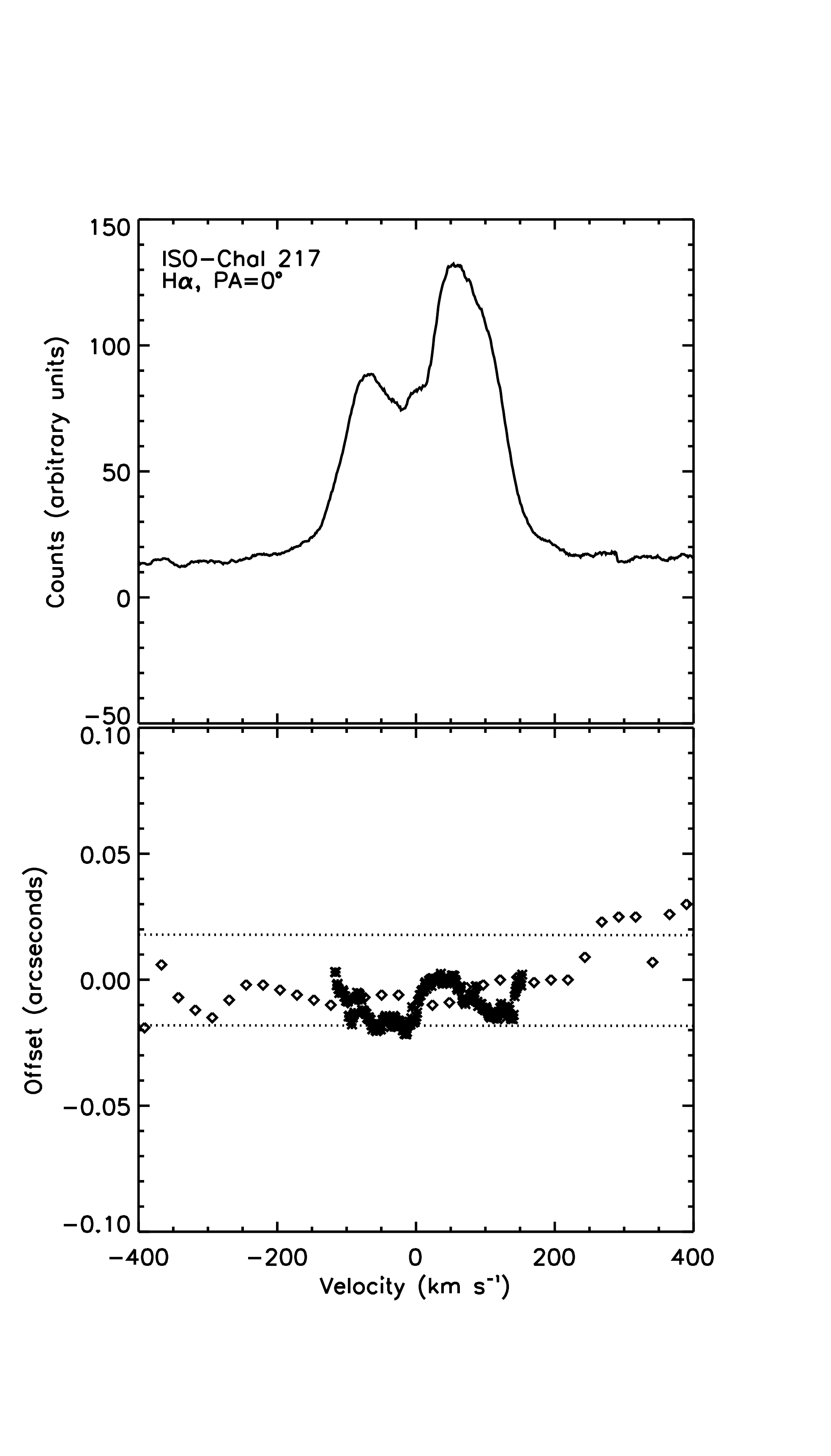}
\includegraphics[width=9.5cm]{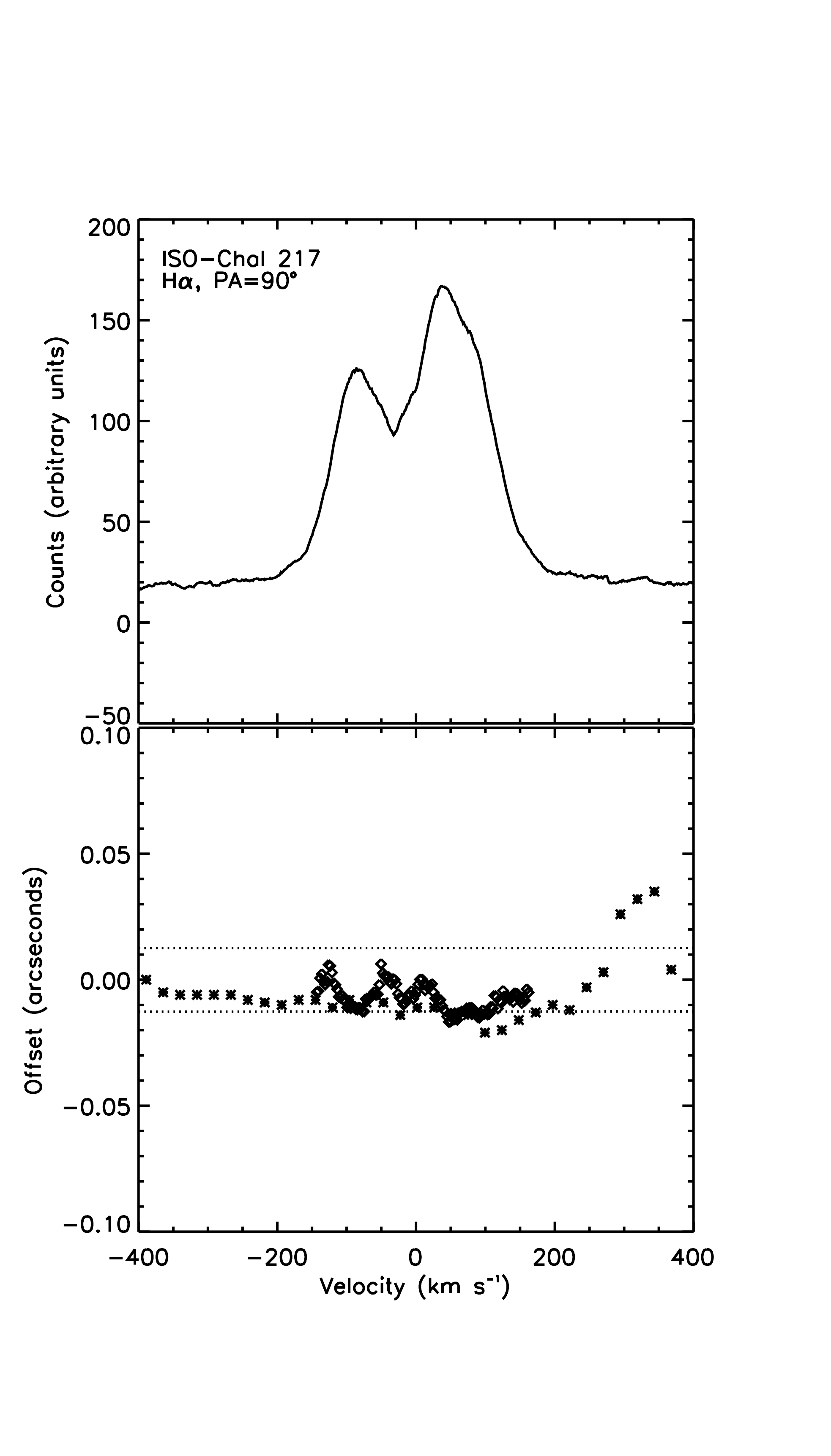}
 \caption{Spectro-astrometric analysis of the ISO-Cha1 217 \Ha\ line. Analysis was carried out in the same manner, as outlined for the \OI\ lines. No signal from the outflow is detected in the \Ha\ line. Emission is coincident with the source hence it is reasonable to assume that the bulk of the emission is coming from the accretion flow. For the \Ha\ line the spectro-astrometric accuracy achieved at the line peak is much greater than in the wings or for the the continuum emission. Hence for all the \Ha\ lines emission is summed so that the error in the line-wing and continuum is comparable. }
 \end{figure}
 \end{center}

\begin{center}
\begin{figure}
\includegraphics[width=9.5cm]{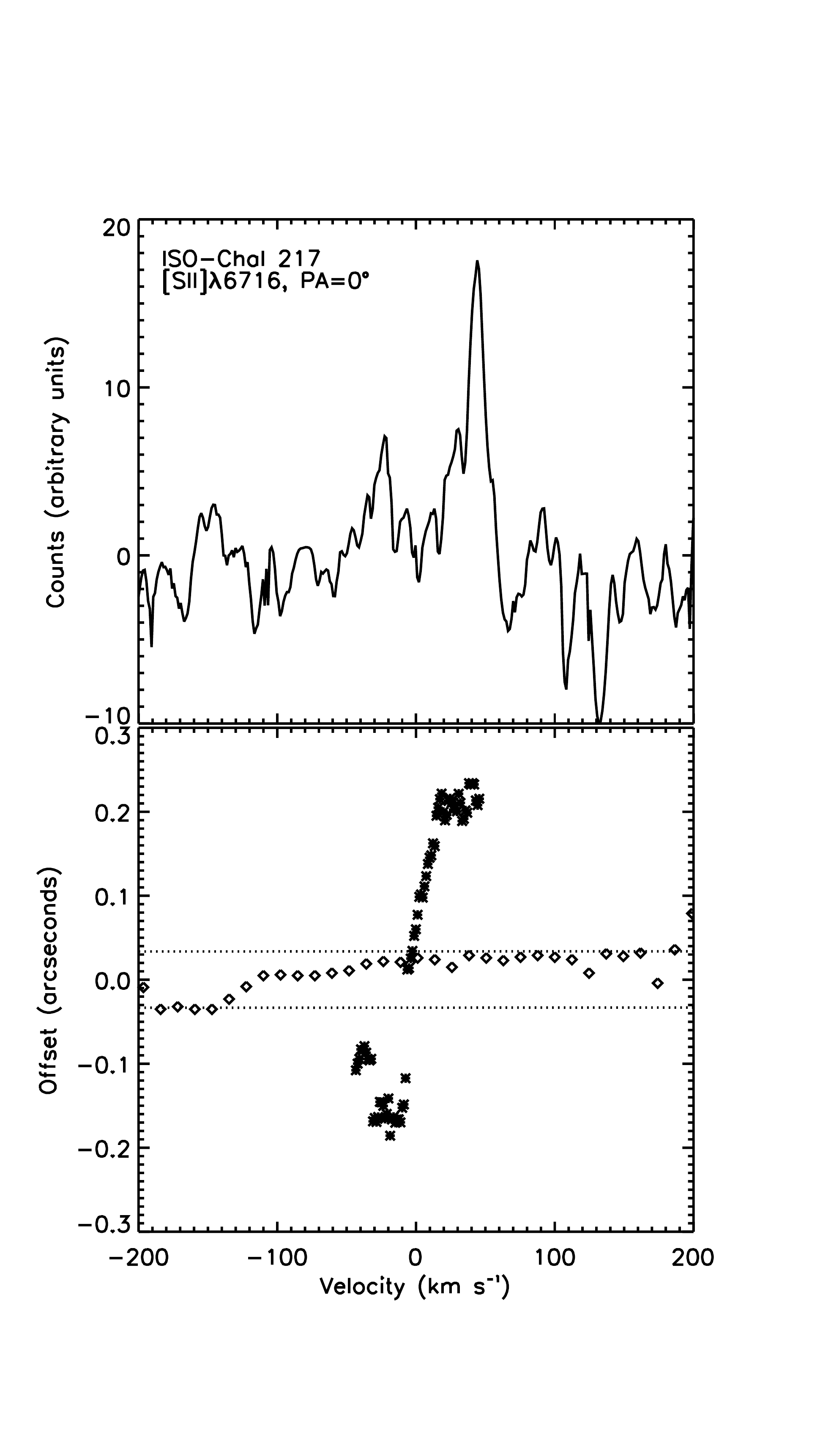}
\includegraphics[width=9.5cm]{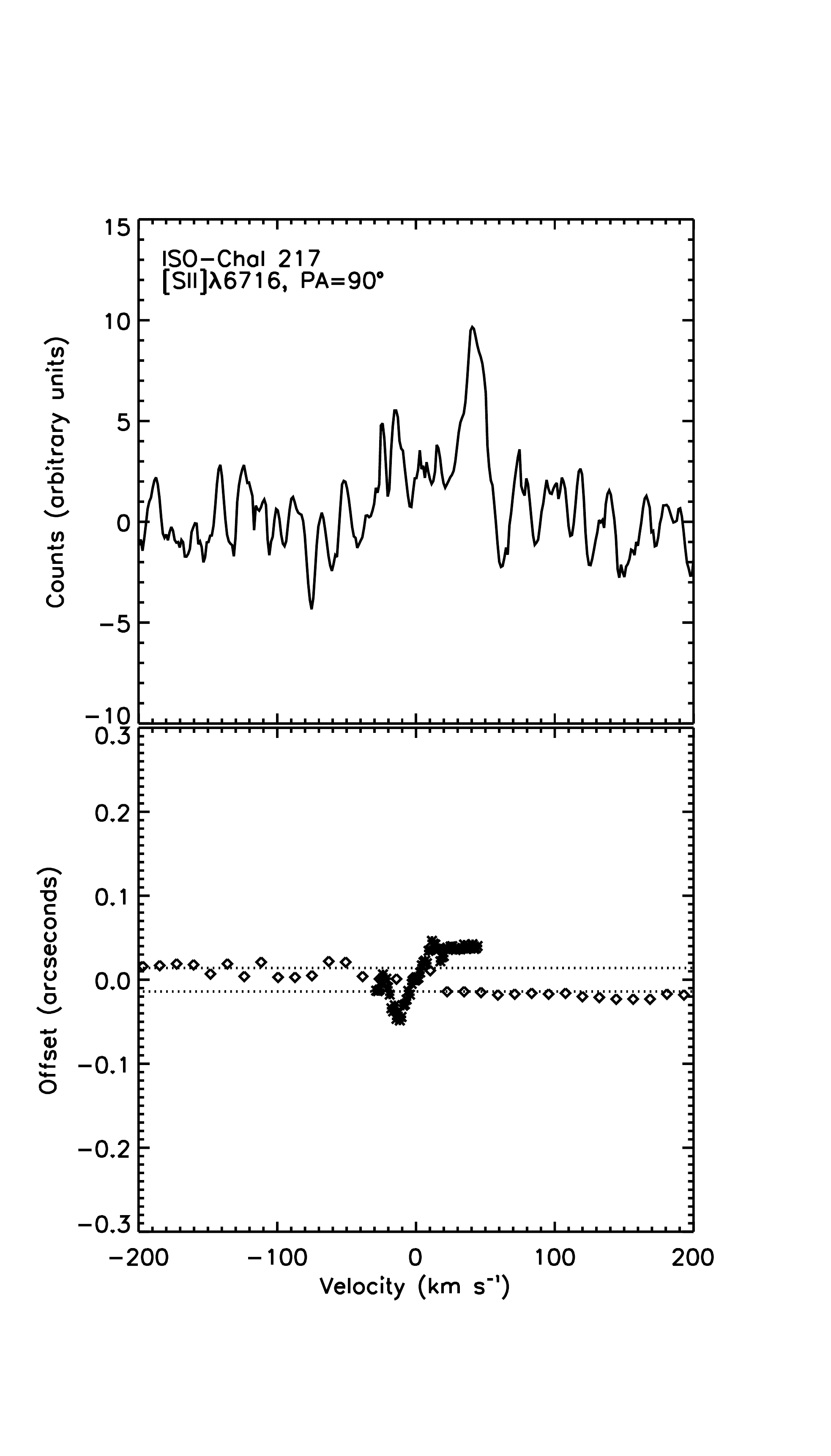}
 \caption{Spectro-astrometric analysis of the ISO-Cha1 217 [SII]$\lambda$6716 line. Again the line was observed at a PA of 0$^{\circ}$ and 90$^{\circ}$ degrees. The blue and red-shifted emission is more clearly separated spectrally than in the \OI\ lines. In addition it is offset further from the BD position. Analysis of the [SII]$\lambda$6716 line constrains the outflow PA at $\sim$  193$^{\circ}$. This is smaller than but similar to the estimate from the \OI\ lines. Note the asymmetry in the red and blue lobes of the outflow in both velocity and intensity. }
 \end{figure}
 \end{center}

\begin{center}
\begin{figure}
\includegraphics[width=9.5cm]{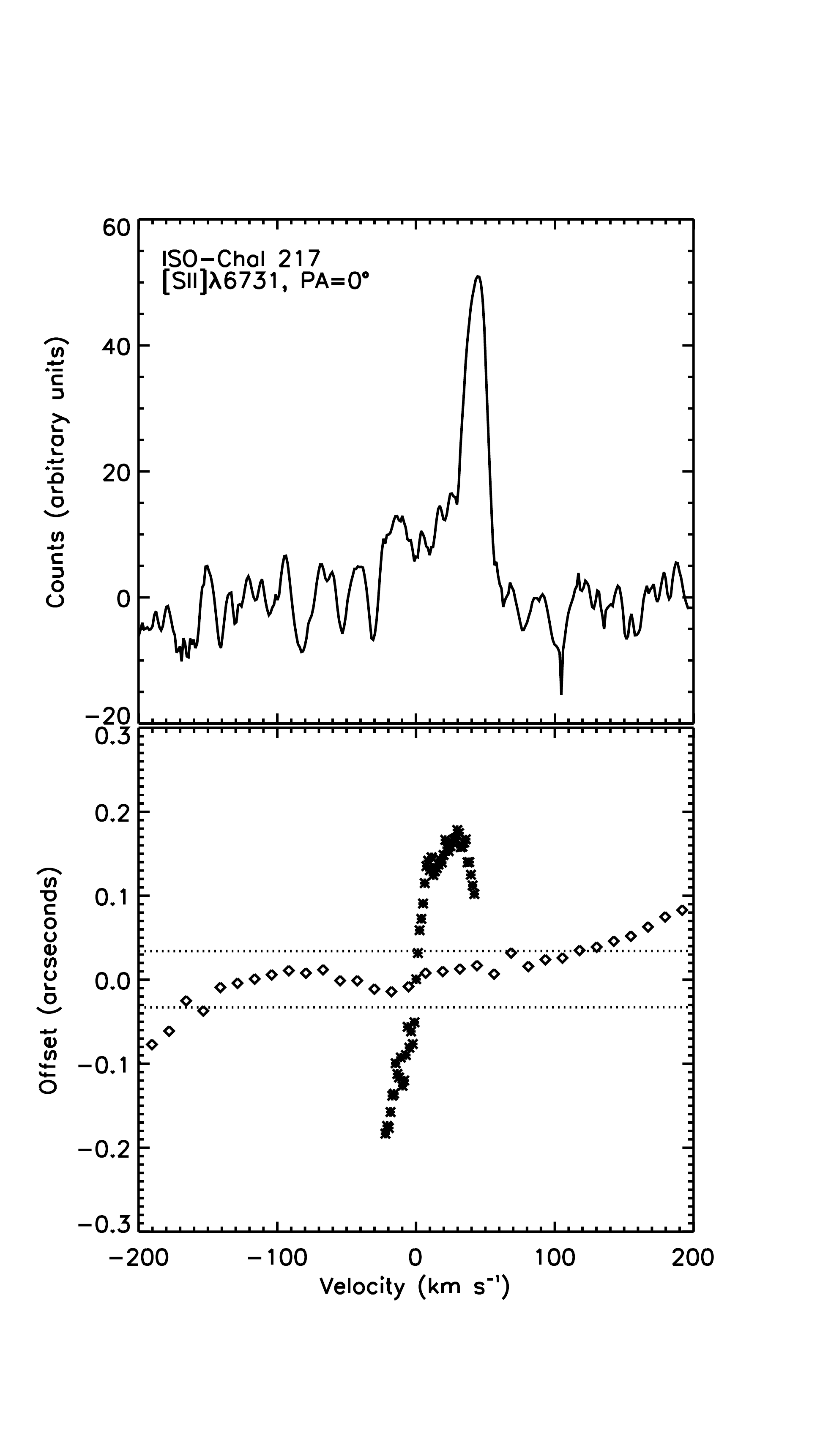}
\includegraphics[width=9.5cm]{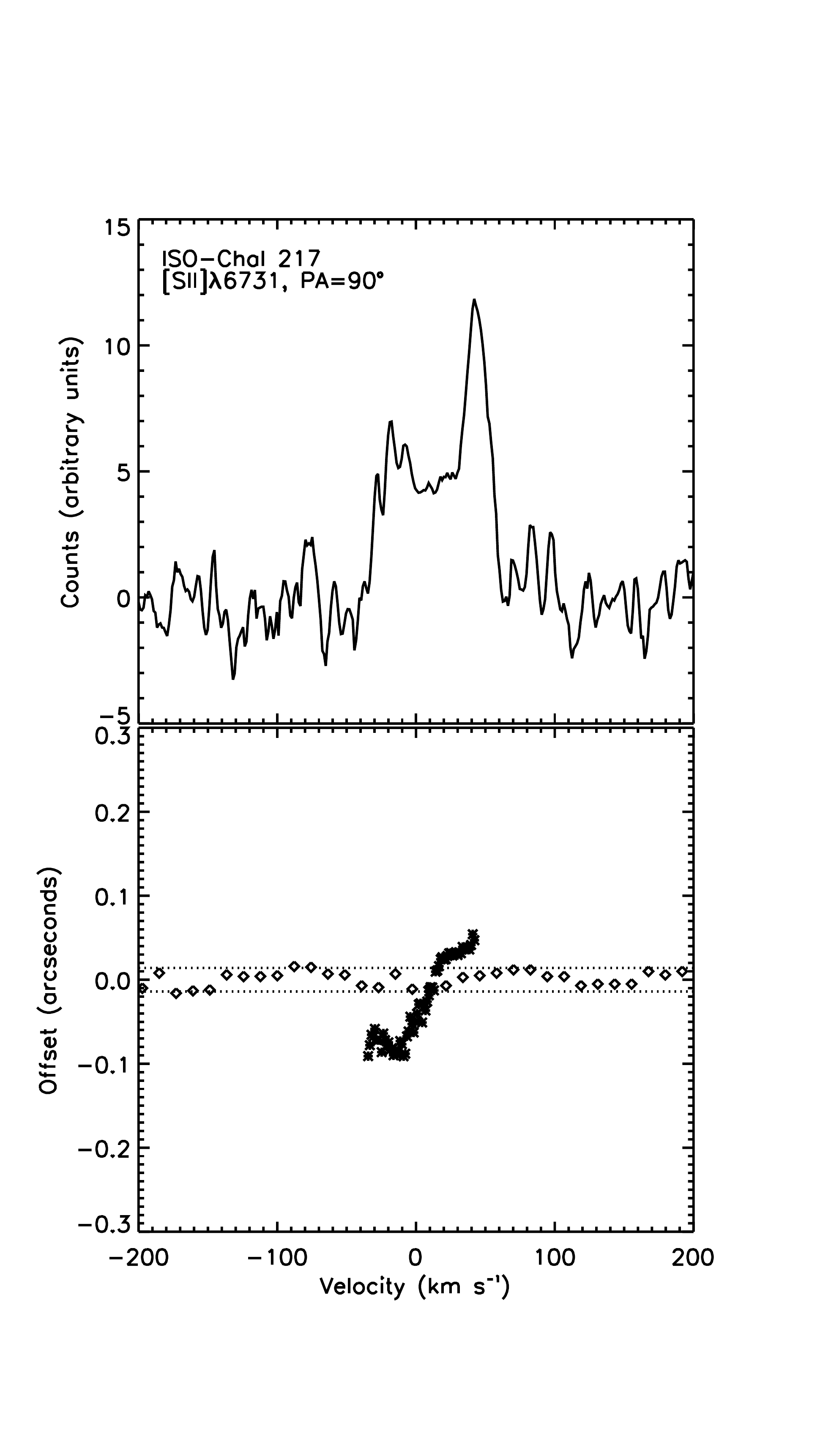}
 \caption{Spectro-astrometric analysis of the ISO-Cha1 217 [SII]$\lambda$6731 line. Analysis agrees with results for the [SII]$\lambda$6716 line and constrains the outflow PA at $\sim$ 200$^{\circ}$. Again the red and blue lobes of the outflow are clearly separated and the asymmetry is especially clear. The difference between the intensity of the red and blue lobes as traced by the [SII]$\lambda$6731 line at 0$^{\circ}$ is striking.}
 \end{figure}
 \end{center}

\begin{center}
\begin{figure}
\includegraphics[width=11.5cm]{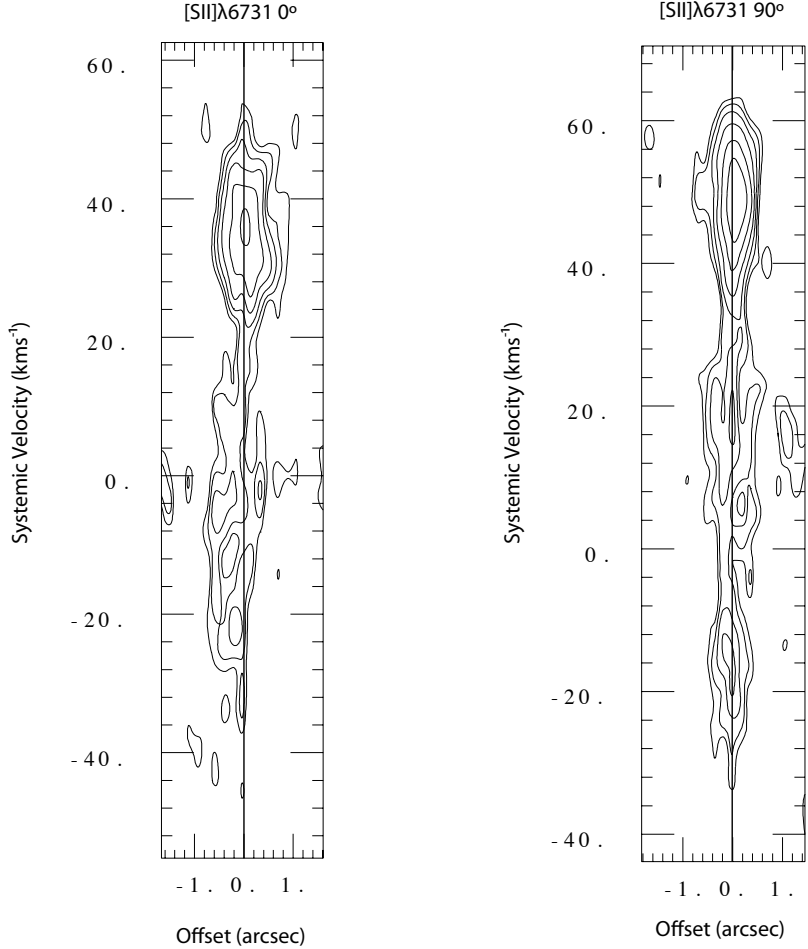}
 \caption{Position velocity diagrams of the \ISO\ [SII]$\lambda$6731 line region, comparing the line at slit PAs of 0$^{\circ}$ (left) and at 90$^{\circ}$ (right). For the PV diagrams the emission line region was smoothed using an elliptical Gaussian filter as described above. Contours begin at 3 times the r.m.s noise and increase by factors of $\sqrt{2}$. The black line delinates the BD position. The PV plots clearly support the spectro-astrometric results. The relative offset between the red and blue-shifted part of the line can be seen by eye in the plot for 0$^{\circ}$. As this offset is a factor of 4 smaller at 90$^{\circ}$ it is not obvious in the corresponding PV diagram.
A difference in brightness between the red-shifted components is also apparent with the red-shifted component at 90$^{\circ}$ appearing marginally brighter. As we are only detecting emission very close to ISO-Cha1 217 and given that our slit width is 1 \arcsec we would not expect to see any significant variation in brightness with PA. The apparent difference here is due to a change in observing conditions between the acquisition of the two spectra with the seeing decreasing from 0\farcs85 for the 0$^{\circ}$ spectrum to 0\farcs7 at 90$^{\circ}$. We measure the integrated flux of the red-shifted components to be comparable with the value being 0.2, 0.45 $\times$ 10$^{-16}$ ergs/s/cm$^{2}$/A and 0.21, 0.5 $\times$ 10$^{-16}$ ergs/s/cm$^2$/A for the red/blue-shifted components at 0/90 degrees respectively.}
\end{figure}
 \end{center}

\begin{center}
\begin{figure}
\includegraphics[width=9.5cm]{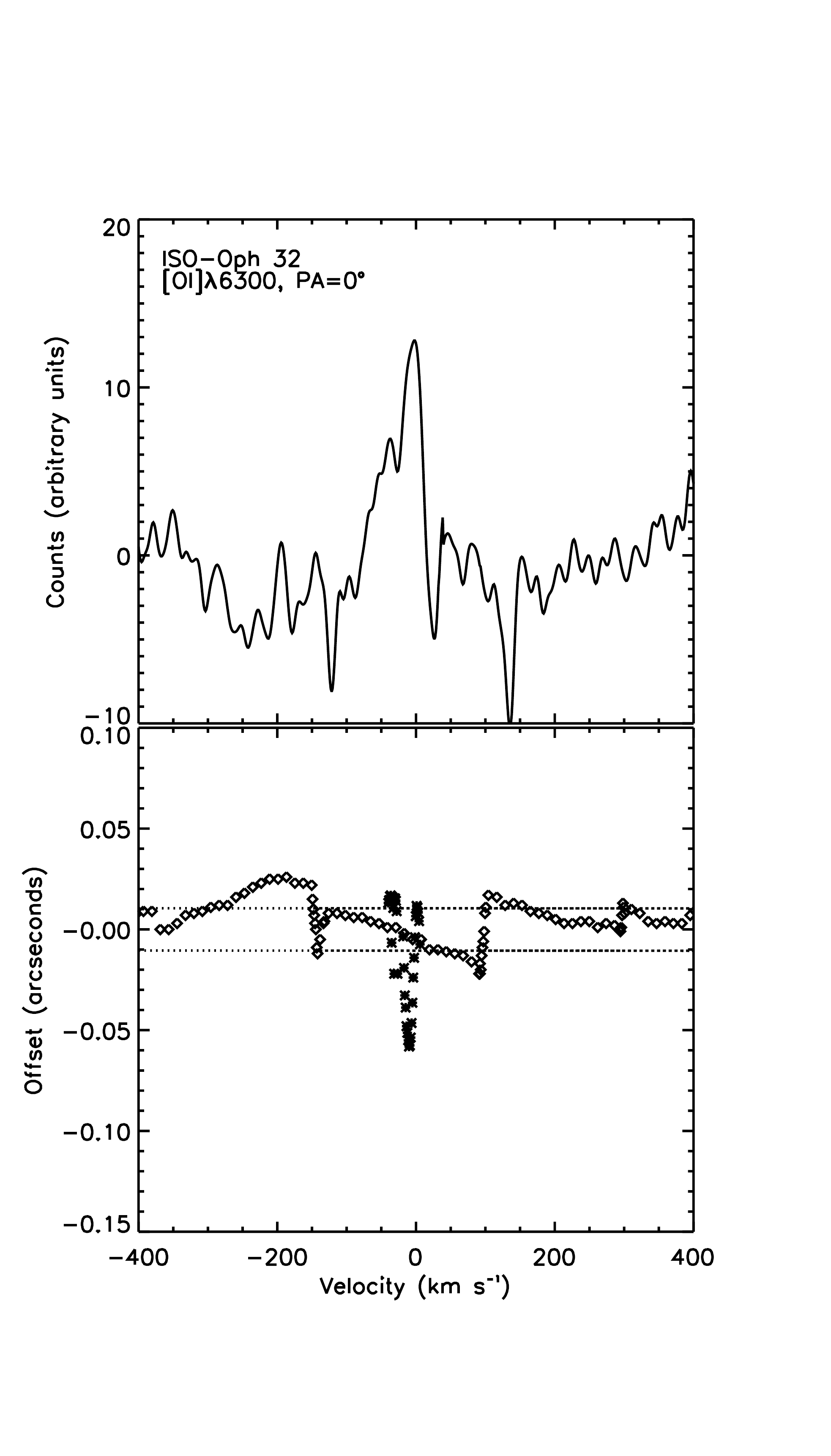}
\includegraphics[width=9.5cm]{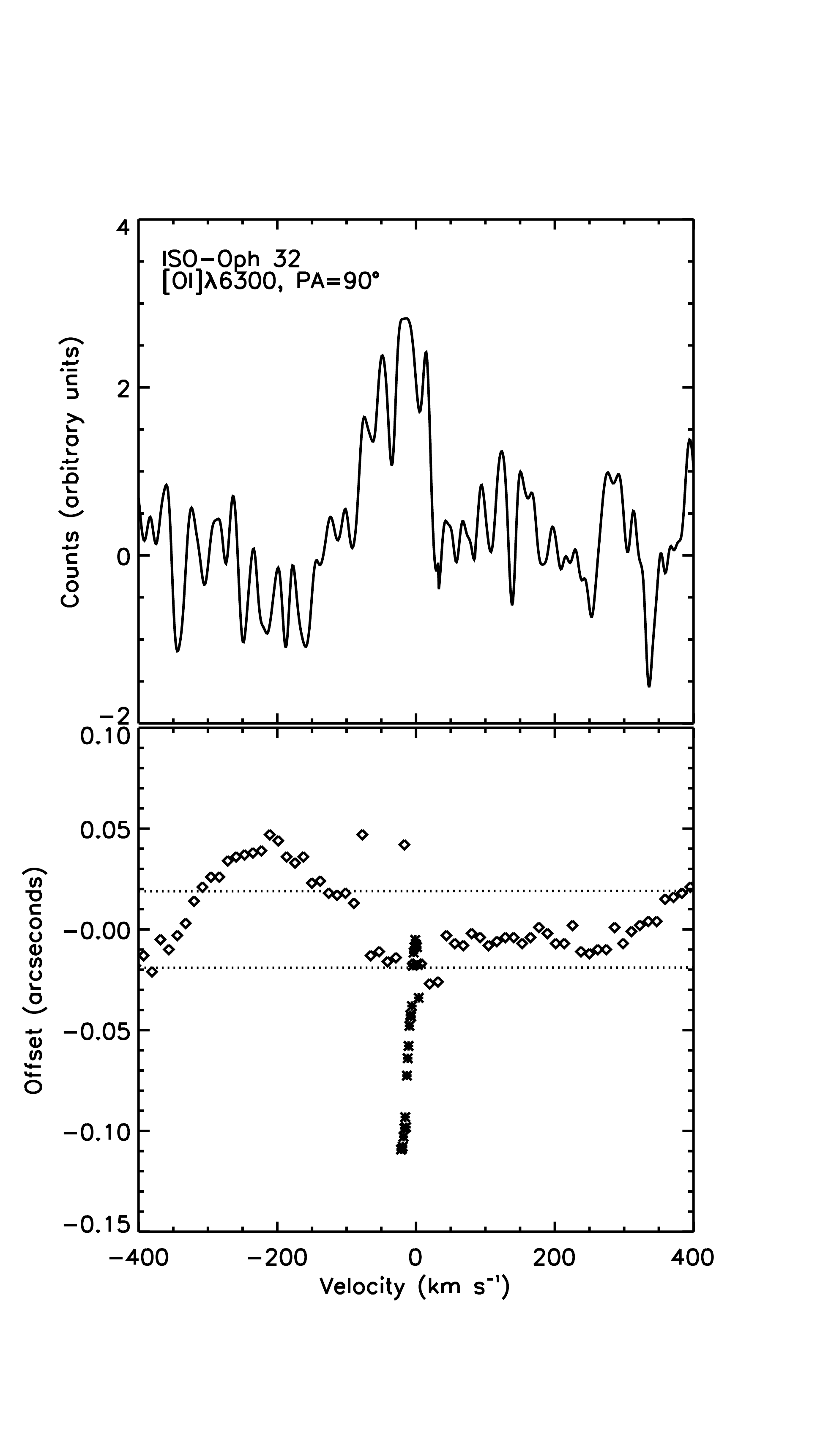}
 \caption{\Ox\ line emission and related spectro-astrometric analysis for the young BD ISO-Oph 32. As reported for other BDs, the \Ox\ line was the only FEL found in the spectrum strong enough for spectro-astrometric analysis. Spectro-astrometry confirms the origin of the line in an outflow. The measured offset in the line is greater at 90$^{\circ}$ than at 0$^{\circ}$ and from this comparison the outflow PA is estimated at 240$^{\circ}$. In the case of ISO-Oph 32 the line and continuum regions were smoothed rather than summed.}
 \end{figure}
 \end{center}
 
  \begin{center}
\begin{figure}
\includegraphics[width=9.5cm]{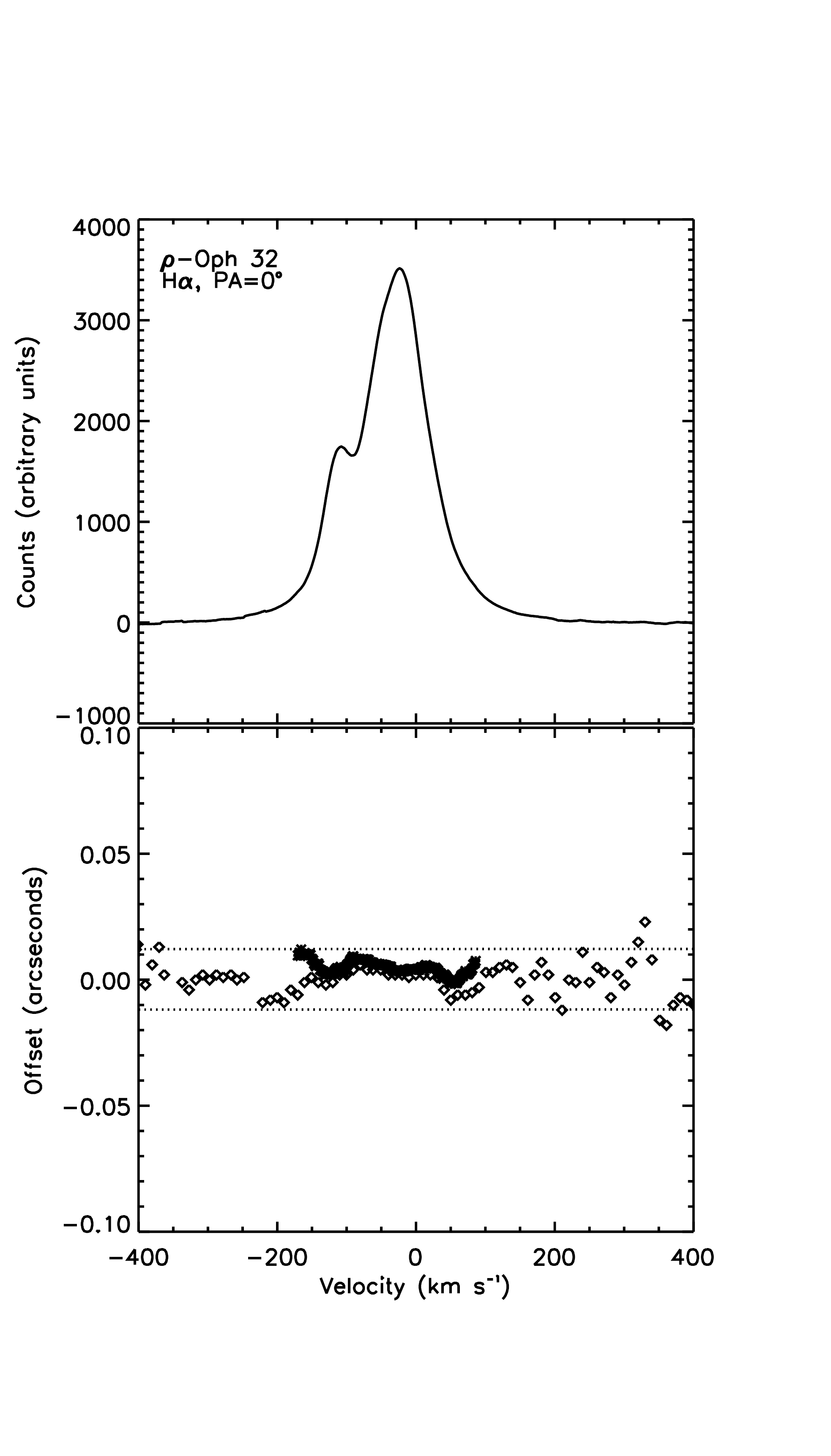}
\includegraphics[width=9.5cm]{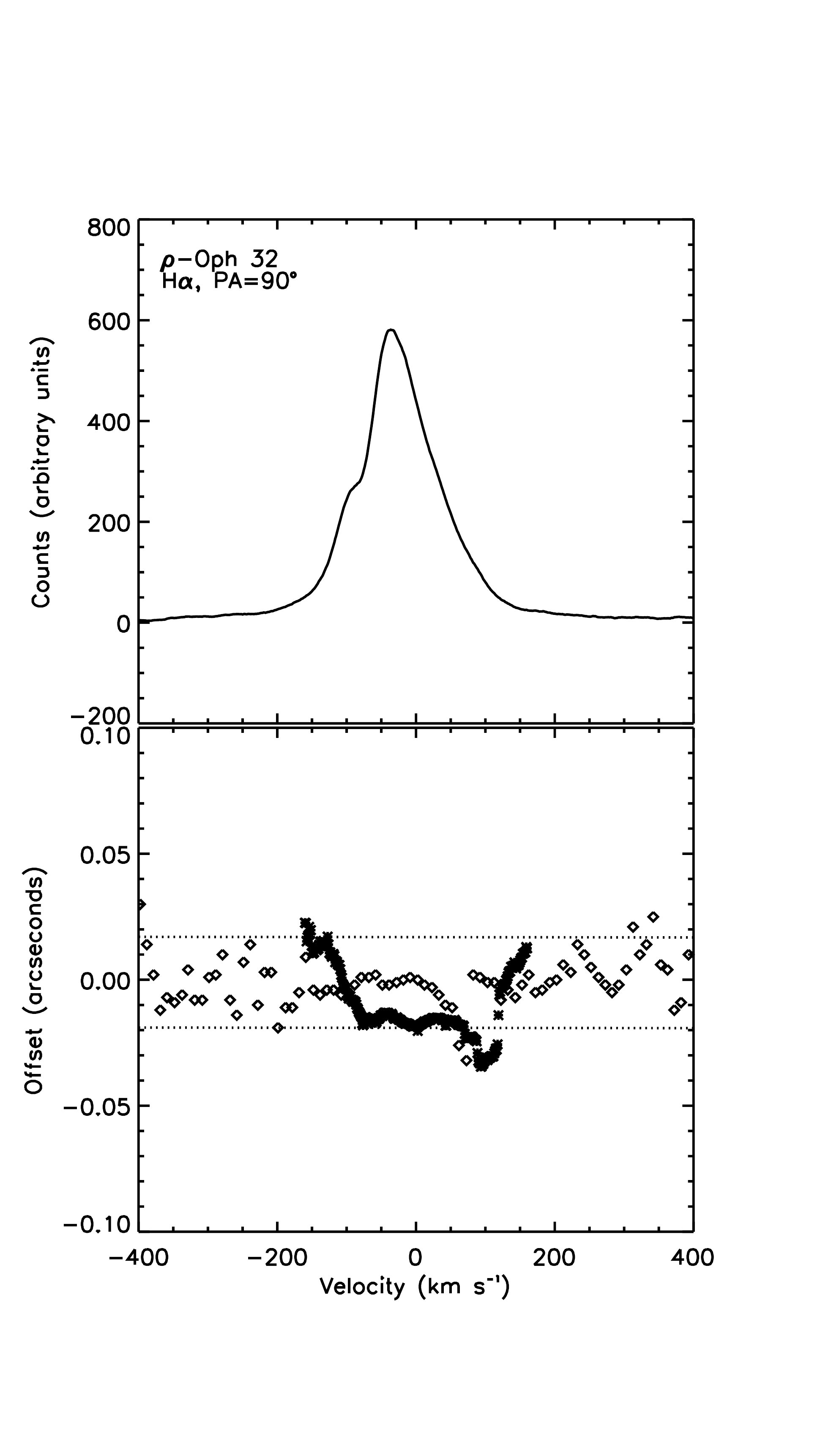}
 \caption{Spectro-astrometry of the ISO-Oph 32 \Ha\ line. As in the case of \ISO\ offsets are a moving sum across the line and continuum position and the line and continuum are summed in such as way that the error in the line-wings and continuum is comparable. The $\pm$ 1-$\sigma$ error is delinated by the dashed lines. As for all the sources discussed in this paper no signature of outflow activity is detected in the \Ha\ line.}
 \end{figure}
 \end{center}

 \begin{center}
\begin{figure}
\includegraphics[width=9.5cm]{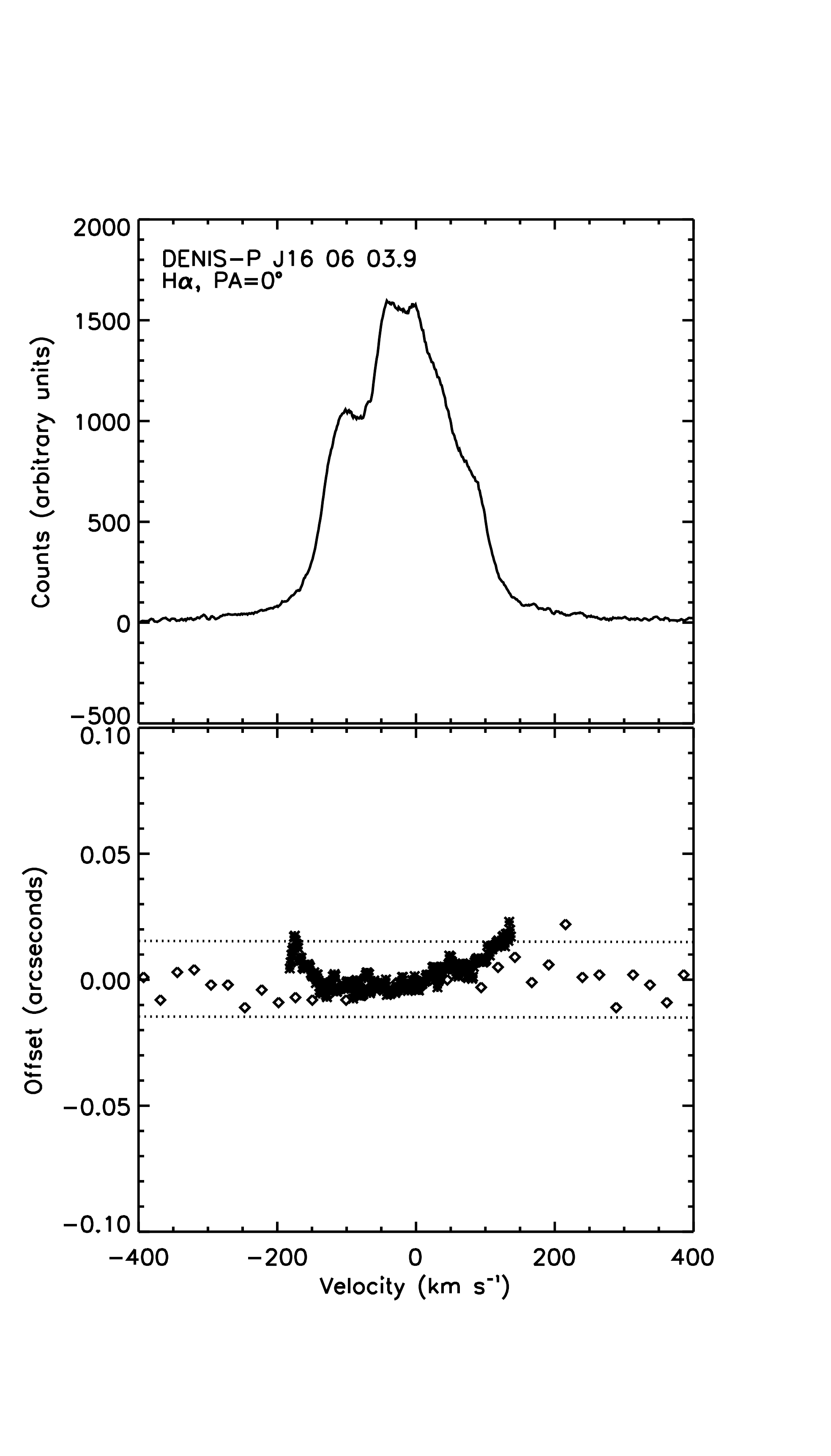}
\includegraphics[width=9.5cm]{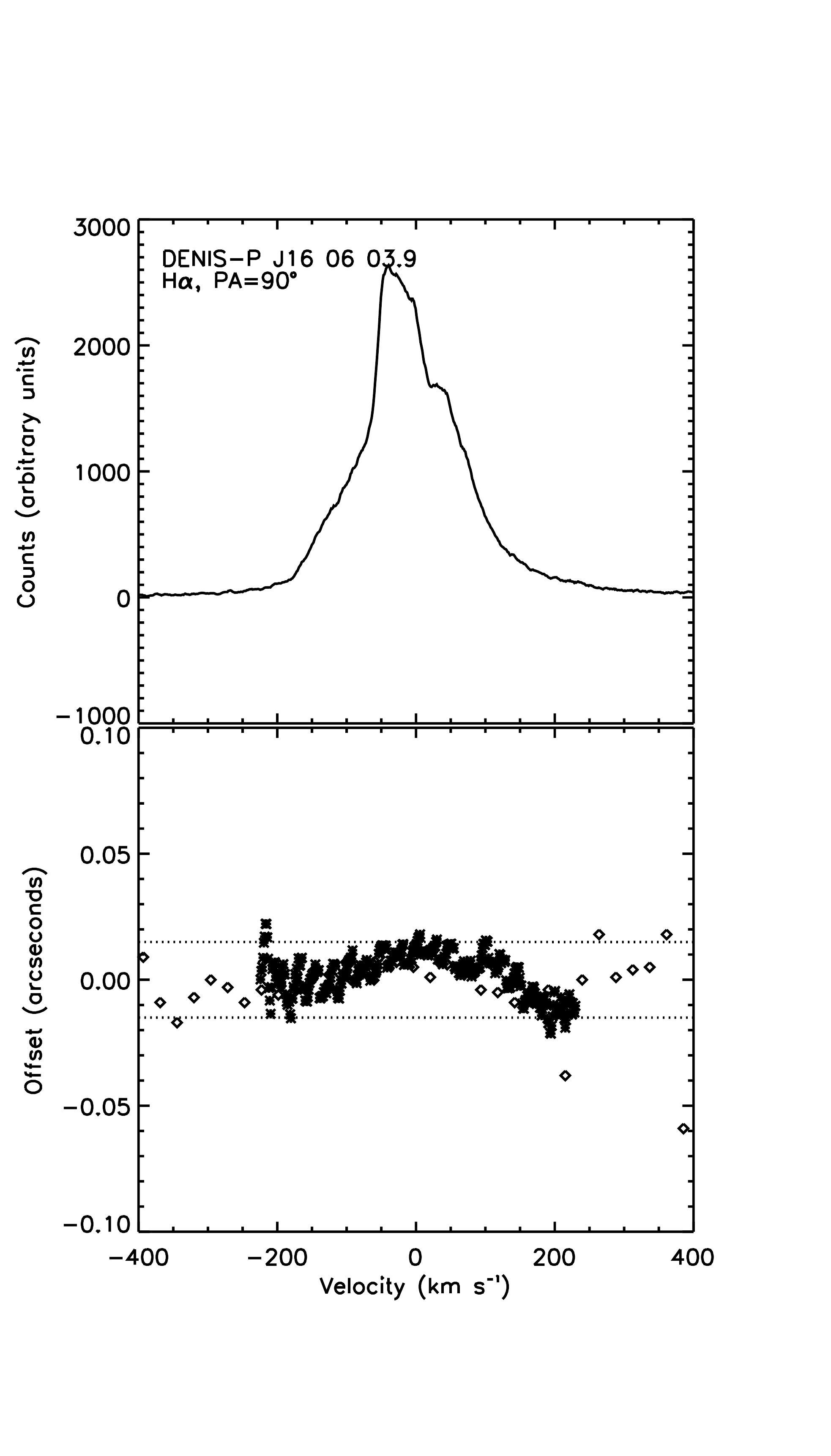}
 \caption{The \Ha\ line and spectro-astrometric analysis for DENIS-P J160603.9-205644. DENIS-P J160603.9-205644 is known to be actively accreting, however no evidence of outflow activity is found in our spectra. As is clear from the analysis shown here the \Ha\ emission is coincident with the source position and therefore must originate very close to the source. This analysis was carried out in the same manner as for all the other \Ha\ lines. }
 \end{figure}
 \end{center}

\acknowledgements{The authors would like to thank A. Natta for her interest in this project and her very valuable assistance with this paper. E.T. Whelan is supported by an IRCSET-Marie Curie International
Mobility Fellowships in Science, Engineering and Technology within 
the 7th European Community Framework Programme. The present work was supported in part by the European Community's Marie Curie Actions - Human Resource and Mobility within the 
JETSET (Jet Simulations, Experiments and Theory) network under contract MRTN-CT-2004 005592). }

\end{document}